\documentclass[usenatbib]{mn2e}
\usepackage{graphicx}
\usepackage{fixltx2e}
\usepackage{textcomp}
\usepackage{aas_macros}
\usepackage{color}
\usepackage{comment}
\usepackage{amsfonts}
\usepackage{amssymb}
\usepackage{amsmath}
\usepackage{epstopdf}

\newcommand{\Mpc}{\rm{\,Mpc}}
\newcommand{\Mvir}{M_{\rm{vir}}}
\newcommand{\Rvir}{R_{\rm{vir}}}
\newcommand{\Msun}{\rm{\,M_\odot}}
\newcommand{\rt}{\tilde{r}}
\newcommand{\rvir}{r_{\rm{vir}}}
\newcommand{\rft}{\tilde{r}_f}
\newcommand{\Mdm}{M_{\rm{DM}}}
\newcommand{\Rcut}{R_{\rm{c}}}
\newcommand{\Mtot}{M_{\rm{tot}}}
\newcommand{\pc}{\rm{\,pc}}
\newcommand{\lcdm}{\Lambda\rm{CDM}}
\newcommand{\cc}{\rm{\,cm^{-3}}}

\newcommand{\fb}{f_{\rm{bar}}}

\newcommand{\adb}[1]{#1} 

\begin{document}
\title{The First Billion Years project: dark matter haloes going from contraction to expansion and back again}
\author[Davis et.~al]{Andrew J. Davis$^1$, Sadegh Khochfar$^{2,1}$, Claudio Dalla Vecchia$^{3,4,1}$ \\
$^1$ Max Planck-Institute for extraterrestrial Physics, PO Box 1312, Giessenbachstr. 1.,  85741 Garching, Germany \\
$^2$ Institute for Astronomy, University of Edinburgh, Royal Observatory, Edinburgh, EH9 3HJ, UK \\
$^3$ Instituto de Astrof\'isica de Canarias, C/ V\'ia L\'actea s/n, 38205 La Laguna, Tenerife, Spain \\
$^4$ Departamento de Astrof\'isica, Universidad de La Laguna, Av. del Astrof\'isico Franciso S\'anchez s/n, 38205 La Laguna, Tenerife, Spain}

\maketitle
  
\begin{abstract}
We study the effect of baryons on the inner dark matter profile of the first galaxies using the \textit{First Billion Years} simulation between $z=16-6$ \adb{before secular evolution sets in}.  Using \adb{a large statistical sample from }two simulations of the same volume and cosmological initial conditions, one with and one without baryons, we are able to directly compare haloes with their baryon-free counterparts, allowing a detailed study of the modifications to the dark matter density profile due to the presence of baryons \adb{during the first billion years of galaxy formation.}.  For each of the $\approx 5000$ haloes in our sample \adb{($3 \times 10^7 \Msun \leq \Mtot \leq 5 \times 10^9 \Msun$)}, we quantify the impact of the baryons using $\eta$, defined as the ratio of dark matter mass enclosed in $100 \pc$ in the baryonic run to its counterpart without baryons.  \adb{During this epoch of rapid growth of galaxies,} we find that many haloes \adb{of these first galaxies} show an enhancement of dark matter in the halo centre compared to the baryon-free simulation, while many others show a deficit.  We find that the mean value of $\eta$ is close to unity, but there is a large dispersion, with a standard deviation of $0.677$.  The enhancement is cyclical in time and tracks the star formation cycle of the galaxy; as gas falls to the centre and forms stars, the dark matter moves in as well.  Supernova feedback then removes the gas, and the dark matter again responds to the changing potential.  We study three physical models relating the motion of baryons to that of the dark matter: adiabatic contraction, dynamical friction, and rapid outflows.  We find that dynamical friction plays only a very minor role, while adiabatic contraction and the rapid outflows due to feedback describe well the enhancement \adb{(or decrement)} of dark matter.  For haloes which show significant decrements of dark matter in the core, we find that to remove the dark matter requires an energy input between $10^{51}$ and $10^{53} \mathrm{\,erg}$.  For our SN feedback proscription, this requires as a lower limit a constant star formation rate between $0.002$ and $0.2 \Msun/\mathrm{yr}$ for the previous $5$ Myr.  \adb{We also find that heating due to reionization is able to prevent the formation of strong cusps for haloes which at $z \sim 12$ have $\leq 10^8 \Msun$.  The lack of a strong cusp in these haloes remains down to $z=6$, the end of our simulation.}  
\end{abstract}
  
\begin{keywords}
cosmology: dark matter -- galaxies: high redshift
\end{keywords}
  
\section{Introduction}
\label{sec:intro}
The currently favoured standard paradigm for structure formation, the $\Lambda$CDM model, assumes that the matter density in the Universe is dominated by dark matter that collapses under the influence of self-gravity to form bound haloes.  Simulations using only dark matter (DM) \citep[e.g.][]{NFW96,NFW97,Navarro04,Maccio07} 
have shown that the DM haloes generally follow a universal (NFW) density profile, 
\begin{equation}
\rho(r) = \frac{\rho_s}{\left(r/r_s\right)\left(1+r/r_s\right)^2},
\label{Eqn:NFW}
\end{equation}
which has an inner profile proportional to $r^{-1}$.  We note, however, that other profiles have been proposed in the literature; \citet{Merritt06} tested seven different profiles and found that the Einasto profile \citep{Einasto89} gave better fits than the NFW profile to their sample of galaxy and cluster size dark matter haloes.  \citet{Navarro04} also found similar results with their halo sample.   Extremely high resolution simulations have shown that inside the radius where the density slope is proportional to $r^{-2}$, the NFW profile underpredicts the density, leading to a steeper, cuspy central profile \citep{Navarro04, Diemand05}; this cuspy behavior is better captured by the Einasto profile. 

Dark matter profiles are subject to change when baryons fall into the halo potential, cool, and collapse to the centre; hence it is not surprising that in contrast to predictions from simulations, observations of dwarf galaxies have shown that a central cusp is not a good fit to the DM density profile \citep{Flores94, Moore94, Burkert95, Swaters03, Kuzio09, Oh11, Salucci12}.  Instead of a central cusp, the DM density flattens to a core-like profile, which can be well fit by a Burkert profile which has an inner slope of zero instead of $-1$ \citep{Burkert95}. Some observed galaxy cluster haloes also show a central flattened profile \citep{ElZant04, Sand08, Newman09}.   Various proposals have been given to explain the difference between the simulations and observations of dark matter haloes of dwarfs and clusters.  

One class of explanations argues against the $\lcdm$ model.  Recently, \citet{Maccio12B} discussed the possibility that warm DM could produce a core profile \citep[see also ][]{Moore94}.  However, they found that the warm DM model which produces the flatter core also severely under-produces the total number of dwarf galaxies.  To address this problem, a mixed cold-warm DM model was then proposed \citep{Maccio12C} to produce a cored profile without significantly affecting the halo mass function down to halo masses of $10^{10} \Msun$.   However, they find that the concentration of the dark matter haloes is only modified for haloes with masses less than $10^{12} \Msun$; thus this model is unable to explain the flattened density profiles of cluster-scale haloes.  \citet{Wechakama11} investigated the possibility that DM self annihilation could produce enough pressure support to explain the dwarf galaxy rotation curves.  However, their canonical model has only a small effect in low-surface brightness dwarf galaxies.  They note that a particular dark matter candidate may be found which can explain the flattened rotation curves with a cuspy dark matter density curve.  However, their model predicts prominent differences between the kinematics of the stars, neutral gas and ionized gas, which have not been observed.
 
Another class of proposals uses interactions between baryons and the DM to flatten the central DM profile.  One possibility for removing a cusp is through rapid outflows of baryons which modify the gravitational potential in the centre of the halo.  As long as the outflow is rapid compared to the local dynamical time, there will be a net energy gain in the DM, even if upon recollapse the potential returns to its initial configuration \citep{Pontzen12,Teyssier13}.  This process has been shown to be effective in simulations at both high redshifts \citep{Mashchenko08} and low redshifts \citep{Governato10, Governato12, Maccio12, DiCintio13} in which the source of the outflow is from supernova feedback.  \citet{Maccio12} also discussed how an offset between the baryonic and DM potential minima can destroy a central cusp.  They showed that feedback can displace the gas such that the location of the baryonic potential well varies significantly with time.  As the gas cycles in and out of the central region of their galaxy, they find that the halo's cusp is flattened to a core profile which matches the observed galaxy sample of \citet{Donato09}.  \citet{Governato12} found that all dwarf haloes in their sample with $10^{10} \Msun < \Mvir < 10^{12} \Msun$ had cored profiles which matched the THINGS \citep{Walter08} and LITTLE THINGS \citep{Hunter12} surveys of nearby field dwarf galaxies.  \citet{DiCintio13} report that the inner profile depends on the stellar to halo mass ratio.  They argue that haloes with few stars can not source the winds required to create a cored profile.  AGN feedback can also source these outflows in larger haloes \citep{Gaibler12, Martizzi12}.  In particular, \citet{Martizzi12} also found that even slow outflows generated during the quiescent AGN phase can cause the DM to adiabatically expand.  

However, a single outflow may not fully provide the solution to the core-cusp problem.  \citet{Ogiya11} found that the potential change due to one baryonic outflow in their simulation was not effective enough to match observations of local dwarf galaxies with flattened profiles. \citet{Governato12} and \citet{Teyssier13} both found that repeated outflows over at least one $\mathrm{Gyr}$ are required to drive a systematic cusp to core transition. \adb{Thus, in our study of galaxy formation during the first billion years, we do not expect to see global systematic core formation as we do not simulate more than one $\mathrm{Gyr}$ and are interested in the physical mechanisms involved during structure formation of the first galaxies.}  

Dynamical friction provides another channel for baryons to transfer energy to the DM.   \citet{ElZant01} have shown that dynamical friction by massive infalling gas clumps can destroy a central DM cusp.  They reported that only $\approx 0.01\% $ of the total halo mass needs to be in the clumps to destroy the cusp; \citet{ElZant04} confirmed this same mass fraction is required to destroy a cusp in galaxy clusters as well.  There are a variety of sources of these gas clumps, including cosmological infall \citep{ElZant04, Tonini06, RomanoDiaz08} and feedback generated clump formation \citep{Maccio12}.  Infalling super massive black holes can also transfer their orbital angular momentum to the DM via dynamical friction \citep{Martizzi12}.  

In addition to baryonic clumps, infalling DM subhaloes may destroy a cusp by dynamical friction \citep{Ma04}.  They find that the ability of a subhalo to destroy a cusp depends on the main halo's merger history, the subhalo concentration, and the total mass fraction in subhaloes.  More concentrated subhaloes and smaller subhalo mass fractions leads to a stronger total decrease in the central density.   Larger mass fractions in subhaloes create a stronger cusp simply due to the increase in total DM mass at the halo centre, and less concentrated subhaloes suffer too much tidal loss before reaching the central regions of the main halo.

Feedback effects on infalling gas will affect its ability  to destroy the cusp.  \citet{Pedrosa10} have shown that in cases with strong supernova feedback, the baryons in the infalling satellites were disrupted at large radii, inhibiting angular momentum transfer in the central regions of the main halo.  They therefore concluded that haloes which are dominated by smooth gas accretion (and hence show disk-like configurations) show more concentration than haloes which show a spheroidal stellar system formed via multiple mergers \citep[see also][]{Pedrosa09}.  \citet{Abadi10} also found similar results; their haloes with active merger histories were less concentrated than haloes with smooth accretion.  However, \citet{Dutton11} came to the opposite conclusion, based on their attempts to simultaneously match the observed slopes and zero-points of the Faber-Jackson and Tully-Fischer relations to their analytic mass models of galaxies and their host dark matter haloes.  They found that early type galaxies favored models with strong halo contraction, while late type galaxies favored models with halo expansion.  Thus, it remains unclear how formation histories affect the final state of the halo.

Studying the inner slope of the dark matter halo requires high spatial resolution in the inner regions of the halo.  This is difficult to obtain in cosmological simulations, unless one zooms in on a few pre-selected haloes \citep[e.g. ][]{Governato12}.   However, a fully cosmological simulation is required so as to fully capture the range of halo formation histories seen in a large statistical sample.  This makes it difficult to simulate a large statistical sample of low redshift dwarf galaxies, due to the trade-off between resolution and computing power.   Thus matching simulations of local dwarf galaxies with observations in large statistical sample is still computationally challenging \adb{and not the focus of this paper.} 

In this work, we study the dark matter haloes of the first galaxies \adb{during the first billion years.}.  With a large statistical sample of high redshift galaxies in a cosmological context, we are able to study which physical processes are most important in modifying the DM density profile \adb{at early times}.
\adb{This is of particular interest given upcoming observations with the James Webb Space Telescope and the European Extremely Large Telescope that plan to study the structure of galaxies with $M_* \sim 10^8 \Msun$ at $z \geq 6$.  Thus, understanding their structure will be critical to interpreting the observations.  Furthermore, these galaxies are forming in a unique epoch.  Haloes are undergoing rapid growth via accretion \citep{Dekel09}, and galaxies show on average rising star formation rates driven by large accretion rates of gas \citep[e.g][]{Finlator11}.  In addition, the merger times are short, leading to many rapid mergers \citep[e.g.][]{Khochfar11}.  With this rapid growth via accretion and mergers, the first billion years of galaxy formation provides an interesting time in which to study the evolution of the central DM density profile.}

Our aim in this paper is threefold. First, we wish to measure the DM density profile of high-redshift haloes hosting the first galaxies, and \adb{determine the impact of baryons by comparing simulations with and without baryons.}  Second, we seek to follow the time evolution of the profile.  Thirdly, in a cosmological context, we study which physical processes drive evolution away from the NFW profile.  The outline of the paper is as follows.   We describe the First Billion Years (FiBY) Project used in our analysis in Section \ref{sec:FiBY}.  We present the results of our simulations in Section \ref{sec:Res}, and discuss the various physical processes which affect the modifications in the DM profile in Section \ref{sec:Disc}.


\section{The FiBY Project}
\label{sec:FiBY}

The FiBY Project is a suite of simulations designed to model the formation of the first stars and galaxies in a cosmological context (Khochfar et al., in preparation; Dalla Vecchia et al., in preparation).  We describe in this section the code and physical models used in FiBY, and refer the interested reader to the above papers for more details.  The FiBY simulation uses the SPH code GADGET-2 \citep{Springel01, Springel05} as modified by the OWLS project \citep{Schaye10}.  We use a cosmological model consistent with recent WMAP results: $\Omega_{\rm{M}} = 0.265$, $\Omega_{\rm{\Lambda}} = 0.735$, $\Omega_{\rm{b}} = 0.0448$, $H_0 = 71 \rm{\,km/s/Mpc}$, and  $\sigma_8 = 0.81$ \citep{Komatsu09}.  The OWLS code includes line cooling for eleven elements (H, He, C, N, O, Ne, Mg, Si, S, Ca, Fe) in photo-ionization equilibrium \citep{Wiersma09a}.  Cooling tables were calculated using CLOUDY v07.02 \citep{Ferland00}.     

Gas particles are converted into a Population II star particle using a pressure law as described in \citet{Schaye08} and follow a Chabrier IMF \citep{Chabrier03} with a mass range of $0.1\,-\,100 \Msun$.  The density threshold for star formation is $n_{\rm{H}} = 10 \cc$ \citep{Maio09}.  Star particles continuously release H, He, and metals into the surrounding gas, with the metal yields based off of models of SNIa, SNII, and AGB stars.  The details of the metallicity feedback and diffusion are given in \citet{Tornatore07} and \citet{Wiersma09b}.  

The FiBY Project implements additional physical models to include star formation in metal free gas.  We include a primordial composition chemical network and molecular cooling for the H$_2$ and HD molecules \citep{Abel97, Galli98, Yoshida06, Maio07}.  We allow for Population III (Pop III) star formation in gas with metallicities less than $Z_{\rm{crit}} = 1.545 \times 10^{-4} \, Z_{\odot}$ \citep{Maio11}, where we assume $Z_{\odot} = 0.013$.  These Pop III star particles return metals back into the surrounding gas with yields from \citet{Heger02} and  \citet{Heger10}.  We use a top heavy Salpeter IMF \citep{Salpeter55} with a mass range of $21$ to $500 \Msun$ \citep{Bromm04, Karlsson08}.  Once the Pop III stars die, their remnants are treated as black hole particles, which grow as described in \citet{Booth09}.  

The FiBY code includes thermal feedback from supernovae, following \citet{DVecchia12}.   In this model, star particles inject enough thermal energy stochastically to their neighbors to ensure that the radiative cooling time is longer than the local sound-crossing timescale.  With this method, radiative cooling does not need to be turned off to ensure efficient conversion of thermal energy to kinetic outflows. In this model, there are only two free parameters.  First is the fraction of the supernova's energy to inject into the gas, $f_{\mathrm{h}}$, which we set equal to unity.  Thus, for each supernova, $10^{51} \mathrm{\,ergs}$ are injected into the surrounding gas.  The second free parameter is the desired increase in temperature (or energy) so as to ensure a long radiative cooling time.  We use the fiducial value from \citet{DVecchia12} of $\Delta T = 10^{7.5} \mathrm{\,K}$.

Reionization is incorporated in FiBY using the uniform UV background model of \citet{Haardt01}. Prior to $z\approx12$, we use collisional ionization equilibrium rates, and after $z\approx9$ we use photoionization equilibrium rates.  During the epoch of reionization, we gradually change from collisional to photoionization rates.  As suggested in \citet{Nagamine10}, we include self-shielding of dense gas against ionizing radiation.  This shielding occurs for densities of $n_{\rm{H}} \geq 0.01 \cc$ and scales as the inverse square of the density.  Our reionization model is self-consistent with the emissivity from the galaxies in our FiBY simulation as shown in \citet{Paardekooper12}.  

For this work, we used the FiBY\!\_S simulation, which has a box size of $4 \Mpc$ comoving, with $2 \times 684^3$ particles in gas and DM.  This gives a DM particle mass of $m_{\rm{DM}} = 6161.5 \Msun$, and an initial gas particle mass of $m_{\rm{g}} = 1253.6 \Msun$.  The spatial resolution is given by the gravitational softening parameter, $\epsilon = 166 \rm{\,pc/h}$ comoving, which at $z=6$ gives a physical spatial resolution of $33 \rm{\,pc}$.  We evolve the simulation from $z = 127$ down to $z = 6$.  

In conjunction with this simulation, we also ran a DM only simulation, the FiBY\!\_DM.  This simulation used the same initial conditions as the FiBY\!\_S simulation, allowing us to make direct comparison between haloes with and without baryons.  For the sake of clearer distinction between these two simulations, we will refer to the FiBY\!\_DM as DMO and the FiBY\!\_S as FiBY for the remainder of this paper.  The mass of the dark matter particles in the DMO simulation are a factor of $\Omega_{\rm{M}}/(\Omega_{\rm{M}}-\Omega_{\rm{b}})$ more massive than the DM particles in the FiBY simulation.  When comparing the dark matter mass between the DMO and FiBY simulations, we scale the DMO mass down to match the FiBY particle mass.

To identify the haloes in our simulation, we use the SUBFIND algorithm of \citet{Springel01B}, as modified to include baryons in \citet{Dolag09}.  This algorithm first runs a standard FOF search only on the DM particles, with a linking length of $0.2$ times the mean particle separation.  Gas, star, and black hole particles are then assigned to the same FOF halo as their closest DM particle.   The density of each particle in the FOF halo is then estimated using the standard SPH kernel over the nearest 48 neighbors.  Finally, substructures are located within a FOF halo by searching for local over-densities.  These substructures are then checked for gravitational boundedness, including the thermal energy of the gas in the check.  The virial radius for each subhalo is then calculated as the radius which encloses an over-density of $178 \rho_{\rm{crit}}$. 

Once the subhaloes are identified in both the DMO and the FiBY runs, we cross-match the halo catalogues.  This is done by first identifying all subhaloes in the FiBY run at the final snapshot ($z=6$) which have $\Mdm \ge 3 \times 10^7 \Msun$ ($ \ge 5000$ DM particles) and are also the main subhalo in their FOF halo.  We then search the DMO catalogue for the corresponding subhalo, ensuring that the DMO centre is located within $4 \rm{\,kpc}$ (distances are physical unless otherwise noted) of the FiBY subhalo centre, roughly the virial radius of a $10^9 \Msun$ halo at $z=6$.   (In what follows, we use the term 'halo' to refer to a subhalo in this sample, and NOT to its host FOF halo.)  We use the location of the particle with the minimum potential energy as the centre of the halo.  There are three cases where the closest halo in the DMO simulation is slightly further away than this cut.  In these cases, the haloes are in the process of falling into a neighboring FOF halo, and are simply at different positions in the infalling orbit.  On average haloes are separated by $\approx 200 \pc$ between the DMO and FiBY simulations.  

In addition, it is possible that in the FiBY run, the particle with the minimum potential is a baryonic particle, and not a dark matter particle.  This may induce a systematic offset between the centres of the FiBY and DMO haloes.  At $z=6$, $137$ out of $751$ FiBY haloes have a baryonic particle (predominantly stars) at the potential minimum.  For these haloes, we measured the distance to the closest dark matter particle (which also is always the most bound dark matter particle), and find a mean offset of $11.0 \pc$ and a standard deviation of $4.3 \pc$, well within the gravitational softening length at all redshifts.  In addition, our choice of center will impact measurements of enclosed mass in the halo centre.  We have also defined a centre using a shrinking sphere algorithm, and found good agreement with results using the centre of the potential.  In most halos, the two centres matched within one gravitational smoothing length.

In comparing the same haloes from the DMO and FiBY runs, we find that the DMO halo is nearly always more massive than the total mass of its FiBY counterpart.  The difference is of order $5\%$ for haloes with $\Mtot > 10^8 \Msun$ and $10\%$ for less massive haloes.  The mass difference between the FiBY and DMO simulation is due to the fact that the baryons in the FiBY simulation can be heated and pushed out of the halo.  The virial radius in the DMO simulation is also slightly larger than the FiBY halo by $5\%$. 

We then follow the most massive progenitors of each halo back in time until the progenitor has less than $5000$ DM particles, allowing us to track the evolution of the density profile in our haloes.  The merger trees are based on the algorithm in \citet{Springel01} and were generated using code developed by \citet{Neistein12}.  In this method, a halo B at redshift $z_B$ has a progenitor halo A at redshift $z_A > z_B$ if at least half the particles -- including the most bound particle --  in A are in B at $z_B$.  We have a total of $793$ haloes in the FiBY simulation at $z=6$ which meet our mass threshold.  After including all their progenitors which also meet the mass threshold, we have a total of $5217$ haloes in our sample, as well as their counterparts in the DMO simulation.

\section{Results}
\label{sec:Res}

\subsection{Density Profile}

\begin{figure*}
\begin{center}
\includegraphics[scale=0.725]{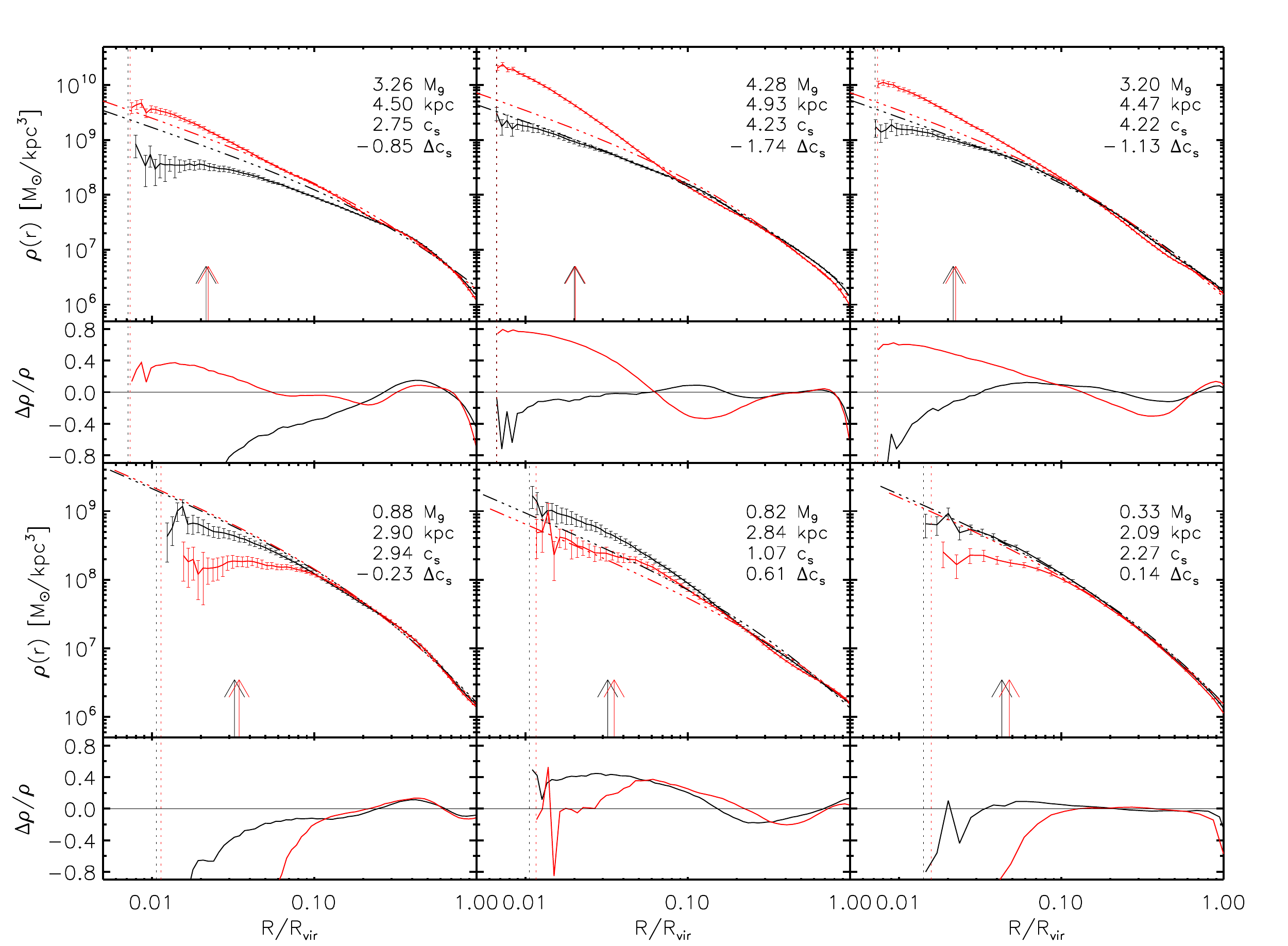}
\caption{Dark matter density profiles for six haloes in our sample at $z=6$.  Black curves mark the profile from the DMO simulation, while the red curve is from the FiBY run.  The profiles are based on a binned histogram and the error bars reflect the Poisson noise in each bin.  The best fitting NFW profile is shown for each run by the dot-dashed curves, and the fractional difference between the simulation and NFW fit is shown in the insets below each panel.   The arrows mark $100 \pc$, the radius we use to define the enhancement of the halo (see Section \ref{ssec:Enh}), and the vertical dashed lines mark the gravitational softening radius of the simulation ($33\pc$ at $z=6$).  The four numbers in each panel show the FiBY total halo mass in units of $10^9 \Msun$, the FiBY virial radius, $\Rvir$, in physical kpc, the concentration of the FiBY NFW fit ($c_s = \Rvir/r_s$), and the difference in concentration between the DMO and the FiBY simulation ($\Delta c = c_{\rm{DMO}}-c_{\rm{FiBY}}$).  We find that some haloes (top row) show significant contraction while others (bottom row) show expansion.  Also note that the modification to the density profiles can extend to $\approx 30\%$ of the virial radius.}
\label{Fig:Mprof}
\end{center}
\end{figure*}

We first compare the density profiles of the FiBY haloes with their counterparts in the DMO simulation.  The profiles are generated using a binned histogram such that each bin has an equal number of particles in it.   We show in Figure \ref{Fig:Mprof} sample density profiles for haloes which show an enhancement of DM in the central regions (top row) and haloes which show a decrement of DM in the central regions (bottom row) compared to their DMO counterpart.  Over-plotted are the best fitting NFW profiles (eq. \ref{Eqn:NFW}) using a Levenberg-Marquardt least-squares fit to the data \citep{Markwardt09} for both the DMO halo and the FiBY halo, as well as the fractional difference between the halo density profile and the NFW profile.  We find that the FiBY haloes may begin to deviate from their DMO counterparts at radii up to $20\%$ of the virial radius, with most halos deviating in the inner $10\%$ of the virial radius.  Thus, the presence of the baryons can affect the dark matter profile to a significant fraction of the virial radius.  In addition, the concentration increases for haloes which show a DM enhancement, as is noted in Figure \ref{Fig:Mprof}; we discuss the change in concentration more in Section \ref{sec:Disc}.  

\adb{We note that the NFW profile generally does a poor job of fitting the DM profile inside $0.1 \Rvir$, often even for the DMO haloes, implying that the NFW may not be the best choice of profile.  In this study, we focus on comparing the DMO to FiBY haloes rather than comparing them to any particular model, which we will do in a follow-up study.  As such, we do not use the fitted profiles for quantifying how the DM profile has changed between the DMO and FiBY simulations.  We only show the NFW profiles in Figure \ref{Fig:Mprof} to guide the eye.}
	
\subsection{Enhancement}
\label{ssec:Enh}
To quantify the change in the DM structure due to the presence of baryons, we define the enhancement, $\eta$, such that
\begin{equation}
\eta \equiv \frac{\Mdm^{\rm{FIBY}}(<\Rcut)}{\Mdm^{\rm{DMO}}(<\Rcut)} \left(\frac{\Omega_m}{\Omega_m-\Omega_b}\right),
\end{equation}
where the masses are the enclosed DM mass inside a cutoff radius, $\Rcut$, for the FiBY and DMO simulations respectively.
The final term scales the mass of the particles in the DMO simulation to that in the FiBY run.  The centre of the halo is defined as the location of the particle with the minimum potential energy.   We use $100 \pc$ as our fiducial cutoff radius, though we have also measured $\eta$ inside $200 \pc$, and find qualitatively similar results; haloes which show enhancement within $100 \pc$ also show it within $200 \pc$, though to a lesser amount.  For the haloes in our redshift and mass range, $100 \pc$ varies between $2\%$ and $20\%$ of the virial radius.  We use a fixed physical radius to better follow the evolution of the halo by ensuring that changes in the evolution of $\eta$ are not simply due to measuring $\eta$ at a larger radius.  For the smallest haloes in our sample ($\Mtot \approx 3\times10^7\Msun$), we have on average $150$ dark matter particles inside $\Rcut = 100 \pc$.  The mass enclosed by $\Rcut$ is usually between $0.5\%$ and $5\%$ of the halo's total mass.  We find that the mass difference between the FiBY and DMO haloes within $\Rcut$ is of order $1\%$ of the DMO halo's virial mass.  

\begin{figure}
\begin{center}
\includegraphics[width = \columnwidth]{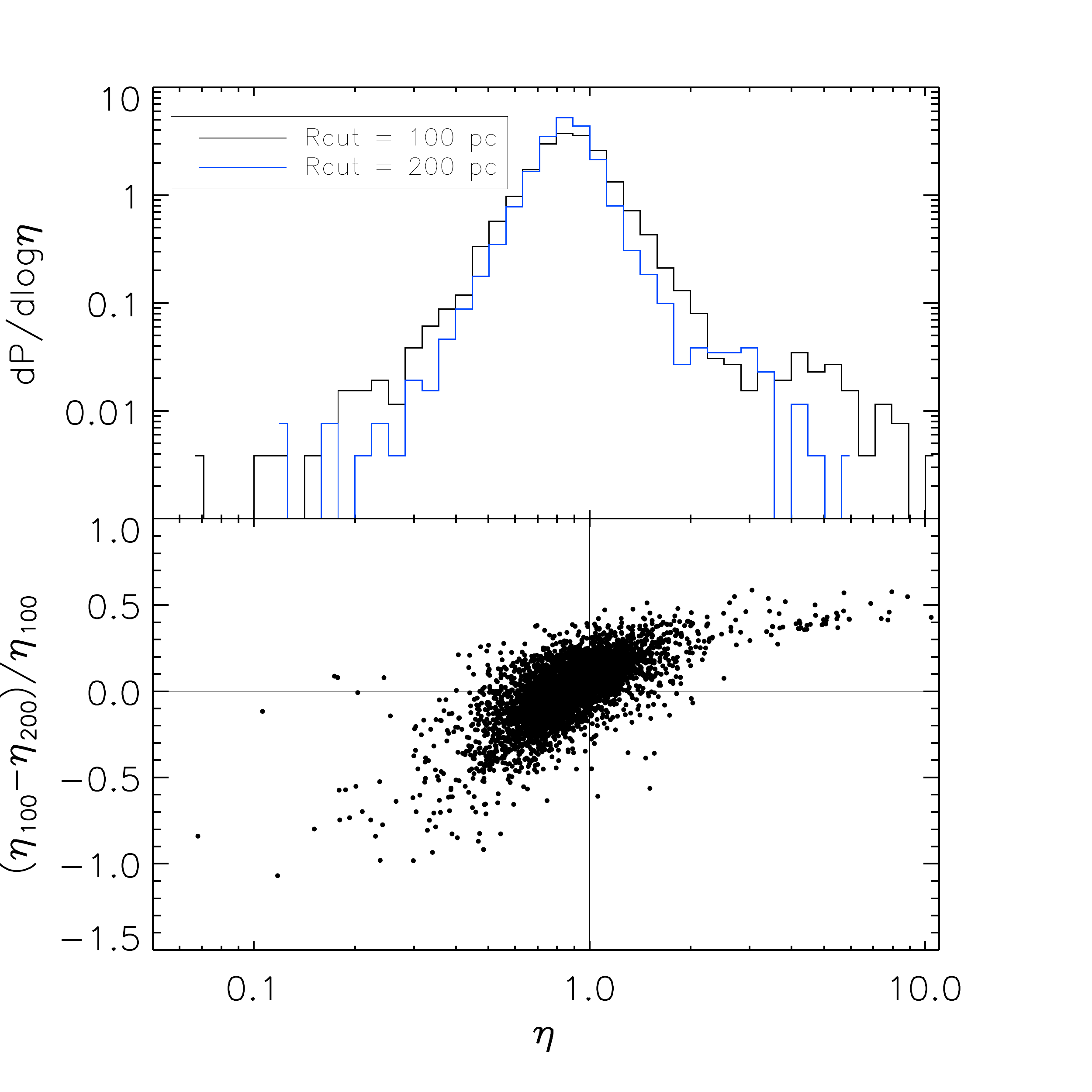}
\caption{Resolution tests to measure $\eta$.  The top panel shows the histogram of $\eta$ measured at $100 \pc$ (black) and at $200 \pc$ (blue).  The distributions are similar, with the $200 \pc$ distribution slightly narrower.  The bottom panel shows the fractional difference in $\eta$ for all haloes in our sample between the two radii as a function of $\eta$ measured at $100\pc$.}
\label{Fig:100_200}
\end{center}
\end{figure}

 In Figure \ref{Fig:100_200}, we show (top panel) the histogram of $\eta$ measured at $100 \pc$ (black) and at $200 \pc$ (blue), as well as (bottom panel) the fractional difference in $\eta$ for all the haloes in our sample.  In the histogram, the high $\eta$ tail is still present at $200 \pc$, just with smaller values of $\eta$.  The width of the distribution is also slightly smaller at $200 \pc$, due to the  smaller modifications to the dark matter profile at larger radii.  

\begin{figure}
\includegraphics[width=\columnwidth, clip=true, trim=15mm 0 0 0]{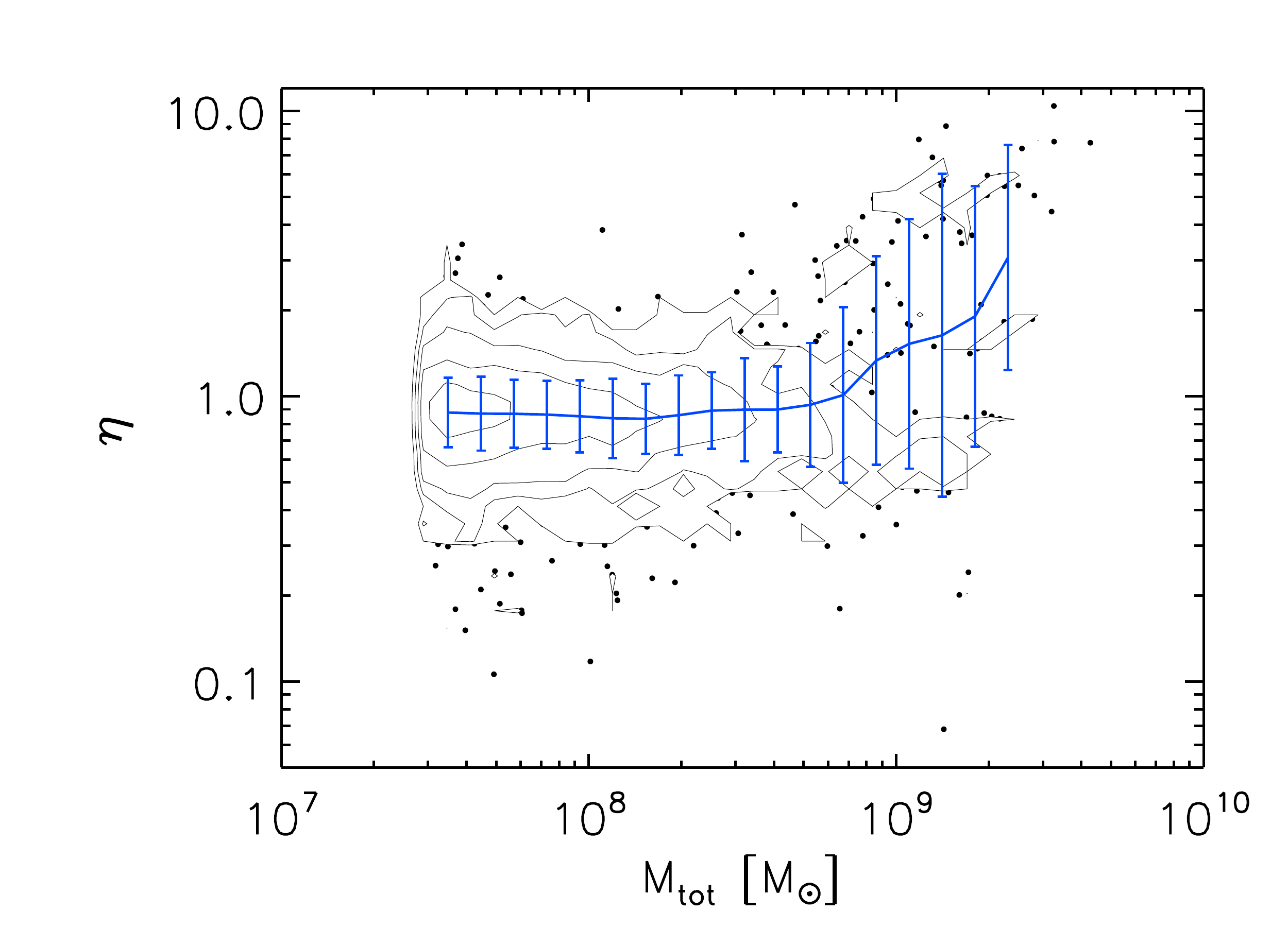}
\caption{Enhancement as a function of total mass for all haloes at all redshifts in our sample.  The blue curve shows the mean trend of $\eta$ with halo mass, and the error bars show the standard deviation of the haloes within each bin.  The points are the haloes which lie outside the lowest contour curve.  We find that there is little evolution of the mean enhancement until the highest masses, which show an offset to large values of $\eta$.}
\label{Fig:Mass}
\end{figure}

We first address trends of $\eta$ with halo mass.  In Figure \ref{Fig:Mass}, we plot $\eta$ versus $\Mtot$ for all the haloes in our sample at all redshifts.  The mean value of $\eta$ for entire distribution is $\bar{\eta} = 0.94$, the median value is $0.87$, and the standard deviation is $\sigma_\eta^2 = 0.495$.  The solid blue curve shows the trend of $\bar{\eta}$ with total halo mass and the error bars denote the $1-\sigma$ deviation in each mass bin.  We find the trend with mass is flat until $\Mtot \approx 6 \times 10^{8} \Msun$, when it rises with increasing halo mass.  The deviation also increases with increasing mass, implying that not all haloes experience the same growth in $\eta$.

\begin{figure}
\begin{center}
\includegraphics[width = \columnwidth]{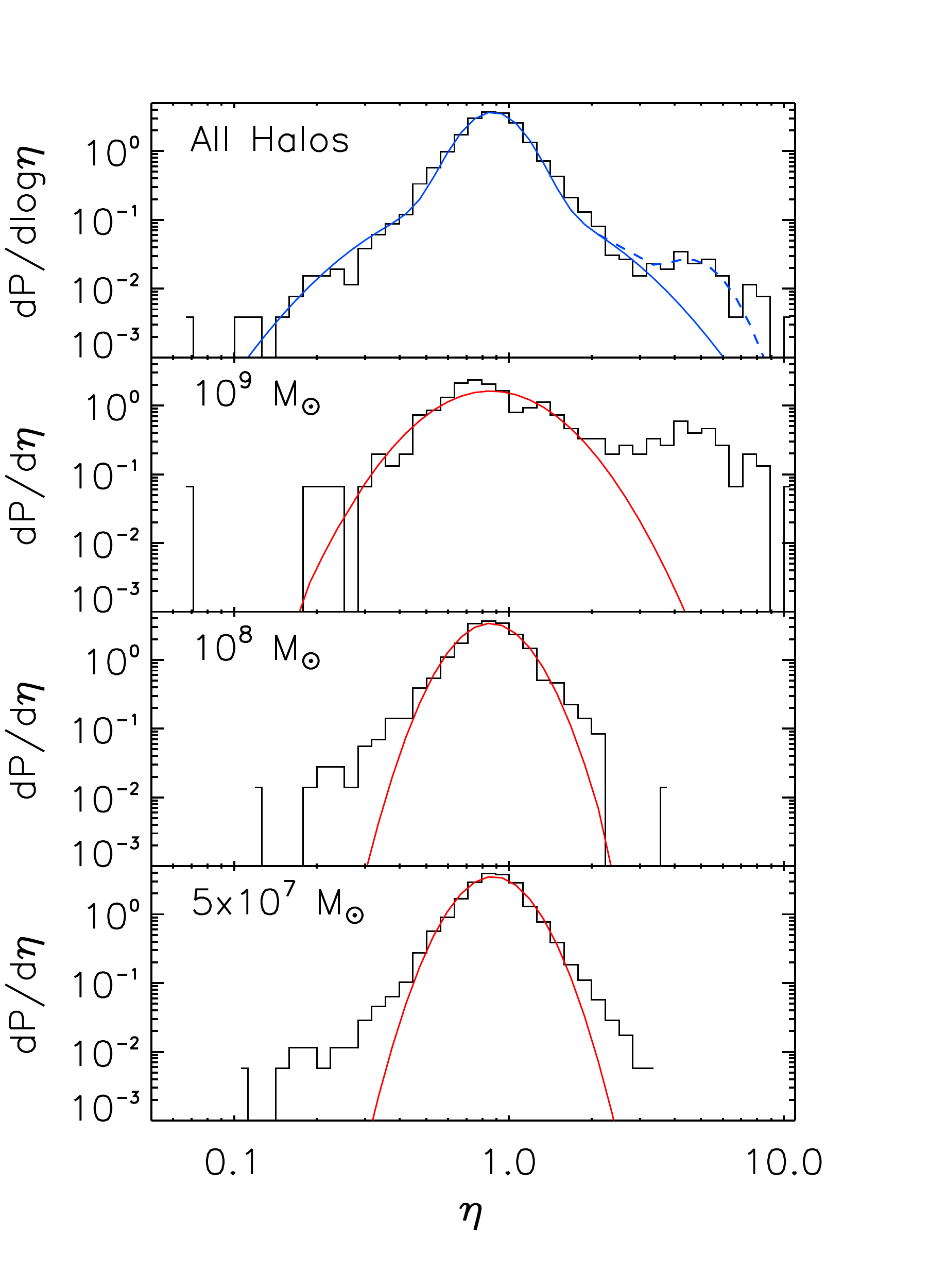}
\caption{Probability distribution of the measured enhancement of the haloes at all redshift slices in our sample (top panel).  We find that the distribution of $\eta$ can be fit with the sum of two log-normal distributions, but it still has a tail at high values of $\eta$.  The tail is fit with a third log-normal shown by the dashed blue curve.  The lower panels give the probability distribution for haloes binned by total halo mass, as well as the best fitting log-normal distribution.  We find that the high mass haloes dominate the high $\eta$ tail of our sample.  See the text for the centres and widths of the log-normal fits.}
\label{Fig:EtaHist}
\end{center}
\end{figure}

In Figure \ref{Fig:EtaHist} we show the probability distribution of $\eta$ for our haloes (top panel).  In blue, we show the sum of two log-normal distributions fit to the distribution.  The central log-normal peaks at $\left<\rm{log}\,\eta\right>_1 = -0.059$ and has a width of $\sigma_1 = 0.134$, while the second log-normal has $\left<\rm{log}\,\eta\right>_2 = -0.085$ and $\sigma_2 = 0.376$.   There is a tail at large values of $\eta$, which is dominated by massive haloes ($\Mtot \approx 10^9$).  We fit the high $\eta$ residuals with another log-normal, denoted in the top panel with a dashed blue curve.  This fit has a center at $\left<\rm{log}\,\eta\right> =  0.67$ and a width of $\sigma = 0.1$.  In the lower three panels, we show the probability distribution of $\eta$ in three mass bins: $\Mtot > 8\times10^8 \Msun$, $8\times10^7\Msun < \Mtot < 3\times10^8 \Msun$, and $3\times10^7 \Msun < M < 8\times10^7\Msun$.  As can be seen, the distribution is roughly log-normal, however at high masses, there is a high $\eta$ tail, while at lower masses, there is a tail at both ends of the distribution.  For the three mass bins (highest to lowest) the mean values of $\eta$ are $\bar{\eta} = (1.54, 0.887, 0.907)$, the median values are $(0.879, 0.855, 0.877)$ and the standard deviations are $1.63, 0.274, 0.264$.

\begin{figure}
\begin{center}
\includegraphics[scale=0.55]{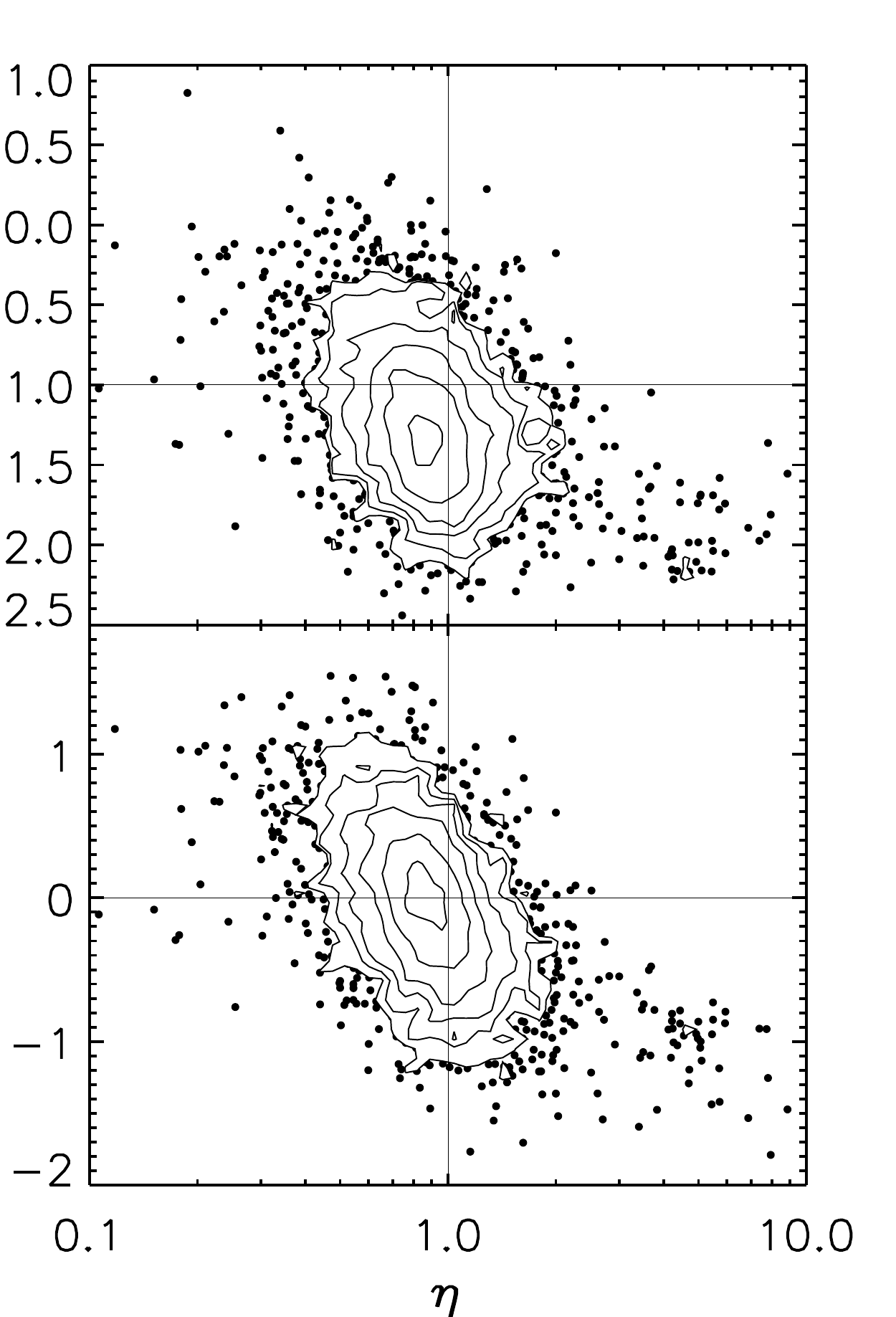}
\caption{\adb{Density profile slope, $\beta = \mathrm{dlog}\rho/\mathrm{dlog}r$, at $\Rcut$ for the FiBY haloes (top) and the difference between the FiBY and DMO runs (bottom) as a function of $\eta$.  The points show the haloes which lie outside the lowest contour curve.  We find that the integral constraint, $\eta$, correlates with the local constraint, $\beta$, reasonably well.  Haloes with large enhancements show steeper slopes than their DMO counterpart.  Haloes showing a decrement of total DM in the core show flatter profiles than their DMO counterparts.}}
\label{Fig:EtaSlope}
\end{center}
\end{figure}

Our definition of $\eta$ is an integral measurement of the density profile inside $\Rcut$.  We can also \adb{use the slope} of the density profile to determine the effect baryons have on the central dark matter profile.   We measure the power law index of the DM profile, $\beta = \mathrm{dlog\,}\rho/\mathrm{dlog\,} r$, at $\Rcut$ by fitting a power law to the density profile between $50$ and $150 \pc$ for both the FiBY and DMO simulations.  We then look at the change in $\beta$ between the two simulations as a function of $\eta$ and show the correlation in Figure \ref{Fig:EtaSlope}.   We find a moderate correlation between $\eta$ and $\beta_{\rm{FiBY}}-\beta_{\rm{DMO}}$, with a Spearman correlation coefficient of $-0.534$.  Haloes with large enhancements show steeper slopes in the core than their DMO counterpart, while haloes showing a decrement have flatter profiles \adb{than their counterpart, albeit with some scatter.  In this paper, we wish to focus on the more robust integrated mass comparison, $\eta$, between the DMO and FiBY haloes.}

\subsection{Cycles of Enhancement}

\begin{figure*}
\begin{center}
\includegraphics[width = \textwidth]{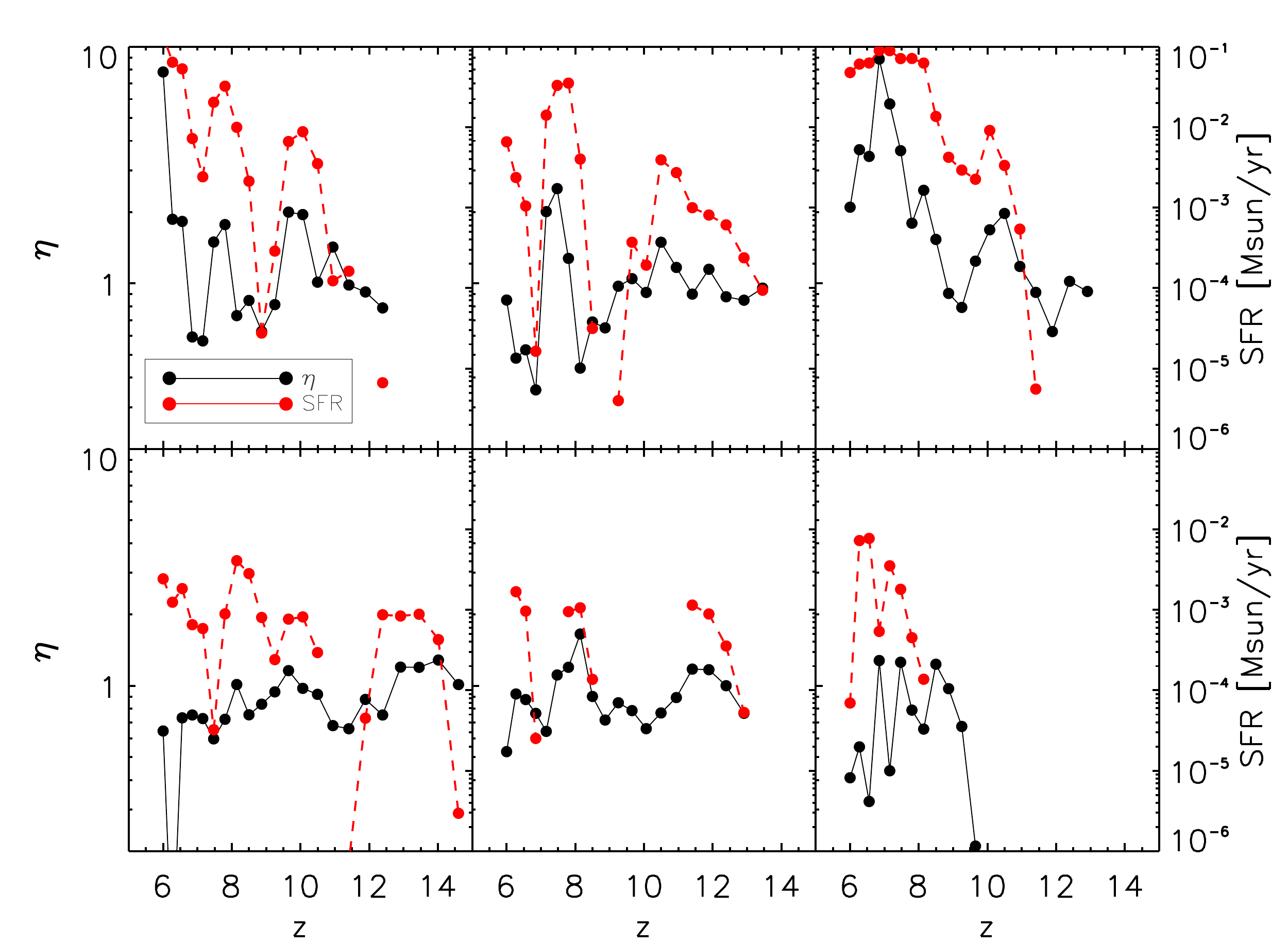}
\caption{Redshift evolution of $\eta$ (black) and star formation rate (red) for six haloes.  We find that the enhancement fluctuates with episodes of star formation.  As cold dense gas falls to the halo centre, a cuspy DM profile develops and stars form.  The stars then input thermal energy into the gas, rapidly blowing out the remaining gas.  The DM responds to the rapid change in potential and expands, thereby decreasing $\eta$.}
\label{Fig:SFREvo}
\end{center}
\end{figure*}

\begin{figure}
\begin{center}
\includegraphics[width = \columnwidth]{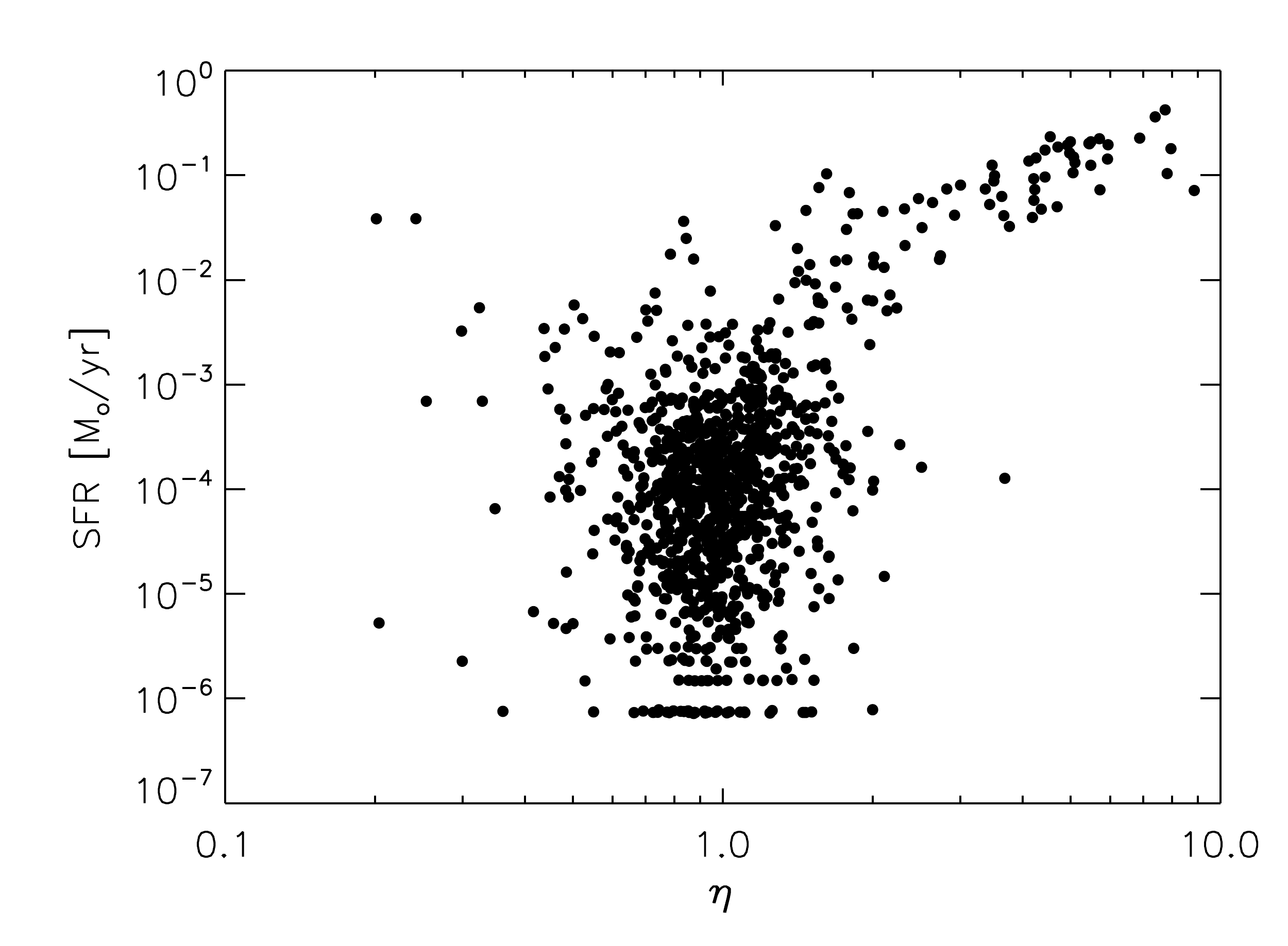}
\caption{Instantaneous star formation rate in the FiBY halo versus $\eta$.  Haloes with no star formation are assigned the value of log(SFR)$= -7$.  We find a strong trend to large $\eta$ with increasing SFR for $\rm{SFR} > 10^{-3} \rm{\Msun/yr}$.  We find that most of the haloes with $\eta < 1$ have little to no star formation -- indicating that these haloes have likely lost their gas due to stellar feedback, which then alters the dark matter structure in the centre of the halo.}
\label{Fig:SFR_eta}
\end{center}
\end{figure}

In addition to having large values of $\eta$, the haloes in the high mass bin of Figure \ref{Fig:EtaHist} also show increased variations in $\eta$ \adb{during the first billion years of their evolution}.  While the overall trend with increasing halo mass is to increasing $\eta$, we find that individual haloes cycle through periods of high and low enhancement.  The change in $\eta$ is largest for the high mass haloes, but lower mass haloes also show similar cycles though of lesser magnitude.  We show in Figure \ref{Fig:SFREvo} the evolution of $\eta$ for six haloes (black curves), showing the cyclical evolution of $\eta$.  Over-plotted in blue is the evolution of the star formation rate (SFR) inside $\Rcut$ for these haloes (blue curves).   We find that peaks in the $\eta$ evolution correlate with peaks in the SFR at the centre of the galaxy.  We show the central SFR versus $\eta$ for all haloes in Figure \ref{Fig:SFR_eta}. Haloes with no central star formation are assigned a value of $10^{-7} \rm{\Msun/yr}$.  We find a clear trend between SFR and $\eta$ for $\rm{SFR} > 10^{-3} \rm{\Msun/yr}$.  

A high value of star formation requires dense gas in the galaxy core.  As this dense gas falls to the galaxy centre, the DM is brought in as well generating an enhancement over the DMO halo counterpart, hence the connection between SFR and $\eta$.  We explore a physical model to explain this enhancement in more detail in Section \ref{Sec:MAC}.  After the gas is converted to stars, feedback heats the gas and expels it from the galaxy centre.  Hence, the star formation also drops.  The drop in SFR  parallels the drop in $\eta$, implying that energy from the feedback is transferred to the DM.  We explore this connection in Section \ref{Sec:BarOut}.  We find that no single star formation episode has the energy to completely shut down star formation permanently in our haloes.  \adb{As gas is accreted freshly onto the growing  halo with high rates \citep{Dekel09} the star formation and $\eta$ will rise again, beginning the enhancement cycle again.  As our study only captures the first billion years of galaxy evolution, we do not see the long term trend to flatter cores seen, for example, in \citet{Governato12} which operate on time-scales longer than our simulation.}

\begin{figure}
\begin{center}
\includegraphics[width = \columnwidth]{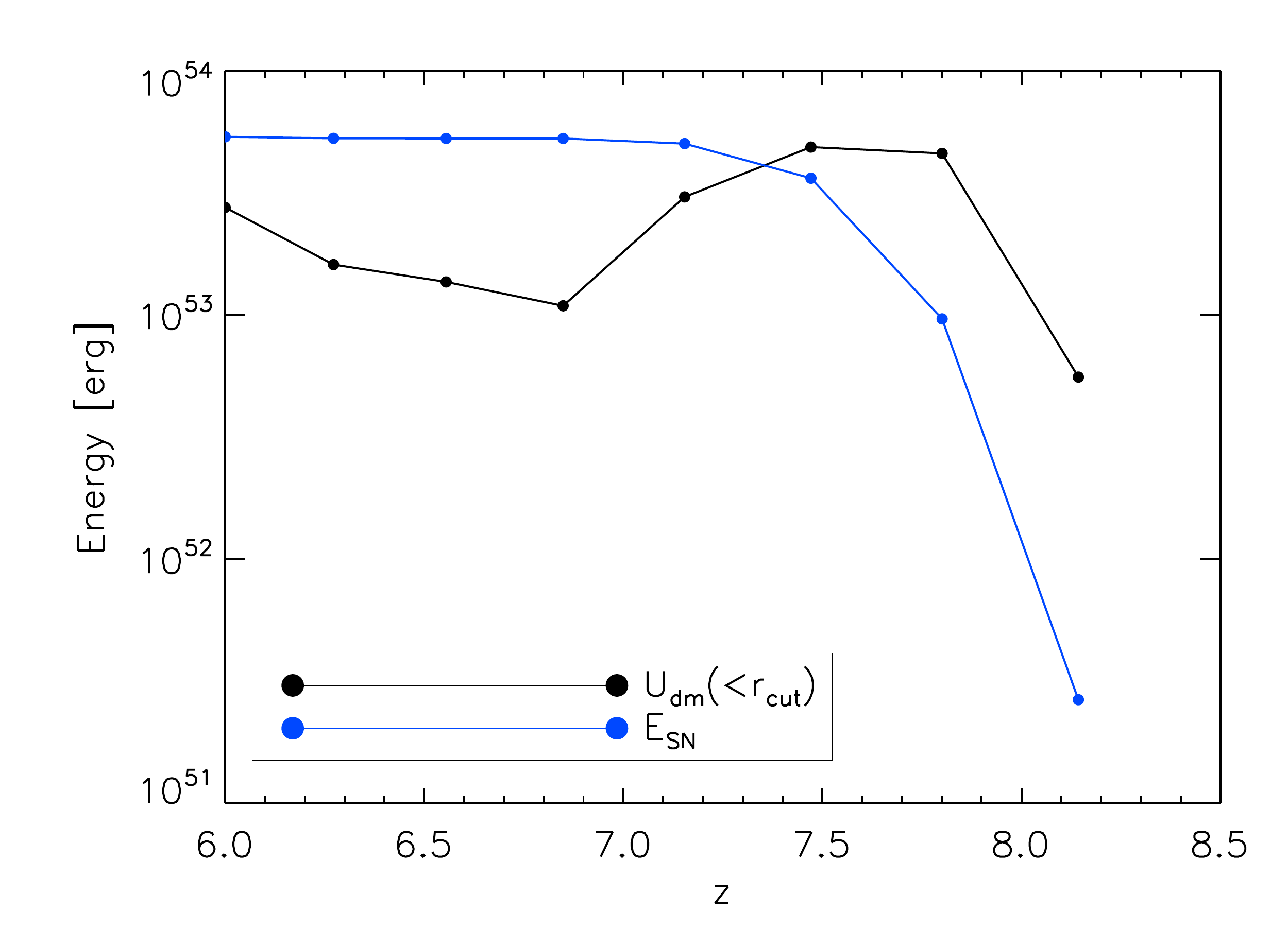}
\caption{Energy evolution of the DM particles inside $\Rcut$ as a function of redshift for one halo as it goes through a cycle of high and low enhancement.  In black, we show the total potential energy of the DM particles inside $\Rcut$.  In blue, we show the cumulative energy injected into the halo from supernovae.   By $z=7.5$, the energy injected by supernovae is comparable to the DM potential energy.  $90 \rm{\,Myr}$ later at $z=6.85$, $\eta$ is at its smallest ($0.36$), the potential is at its shallowest, and the gas is completely evacuated from inside $\Rcut$.}
\label{Fig:H4NRG}
\end{center}
\end{figure}

We selected one cycle to study the state of the gas and energetics of the DM in more detail.  We selected the cycle seen in the top center panel of Figure \ref{Fig:SFREvo} in the redshift range $z=6-8$.  First, we follow the energy in the DM and compare it to the energy injected into the gas by supernova feedback.  In Figure \ref{Fig:H4NRG} we show in black the potential energy, $U_{\rm{DM}}$, of the DM particles inside $\Rcut$ as a function of redshift.  In blue, we show the cumulative energy injected into the gas by supernova feedback, $E_{\rm{SN}}$.  At $z\approx 7.5$, the energy injected by supernovae is comparable to the potential energy of the DM inside $\Rcut$.  This is also the redshift where $\eta$ is at its peak value ($\eta = 2.51$).  After this time, $\eta$ begins to drop and the gas begins to be blown out of the central region of the galaxy.  By $z=6.85$, $\eta$ is at a minimum ($\eta$ = 0.36), only $90 \rm{\,Myr}$ later. 

\begin{figure}
\begin{center}
\includegraphics[width = \columnwidth]{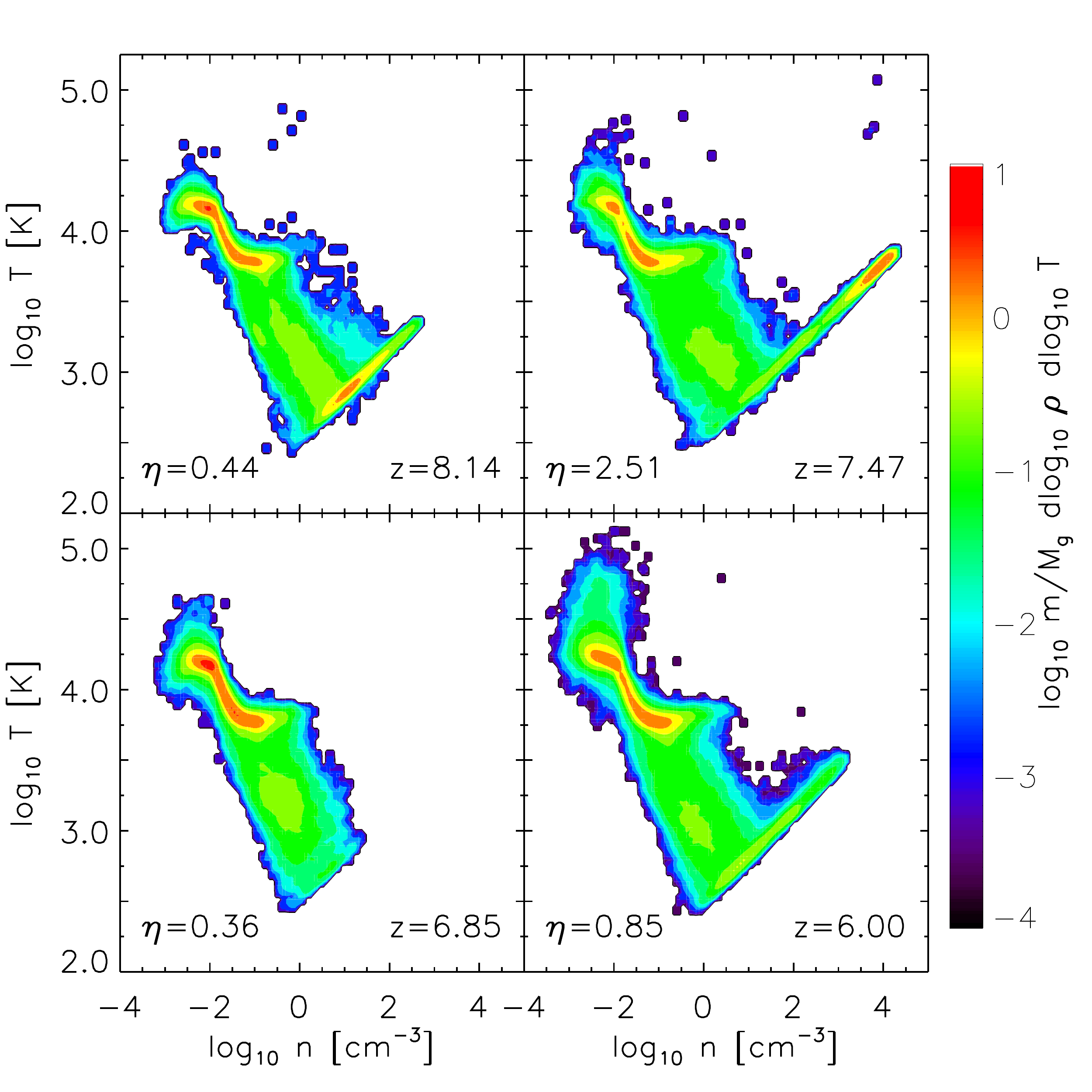}
\caption{Gas phase diagram at four redshifts during one halo's enhancement cycle.  As dense gas is built up between $z=8.14$ and $z=7.47$, $\eta$ increases by a factor of 6.  When the DM shows the strongest enhancement (upper right), there is significant amounts of dense star-forming gas.  When the dense gas is completely removed via supernova feedback, the DM shows the strongest decrement (bottom left).}
\label{Fig:H4PH}
\end{center}
\end{figure}

In Figure \ref{Fig:H4PH}, we show the phase diagram of the gas in the halo at four redshifts during this cycle.  At the start of the cycle ($z=8.14$, upper right panel), $\eta = 0.44$ and the gas is only just denser than the star formation threshold of $n = 10 \rm{\,cm}^{-3}$.  However, by $z=7.47$ (upper left), the fraction of gas above the star formation threshold is dramatically increased and the peak density has increased to $10^4 \rm{\,cm}^{-3}$.  However, by $z=6.85$, the stellar feedback has re-heated the gas, leaving virtually no gas above the star formation threshold, and a very small value of $\eta$.  

\begin{figure}
\begin{center}
\includegraphics[width=\columnwidth]{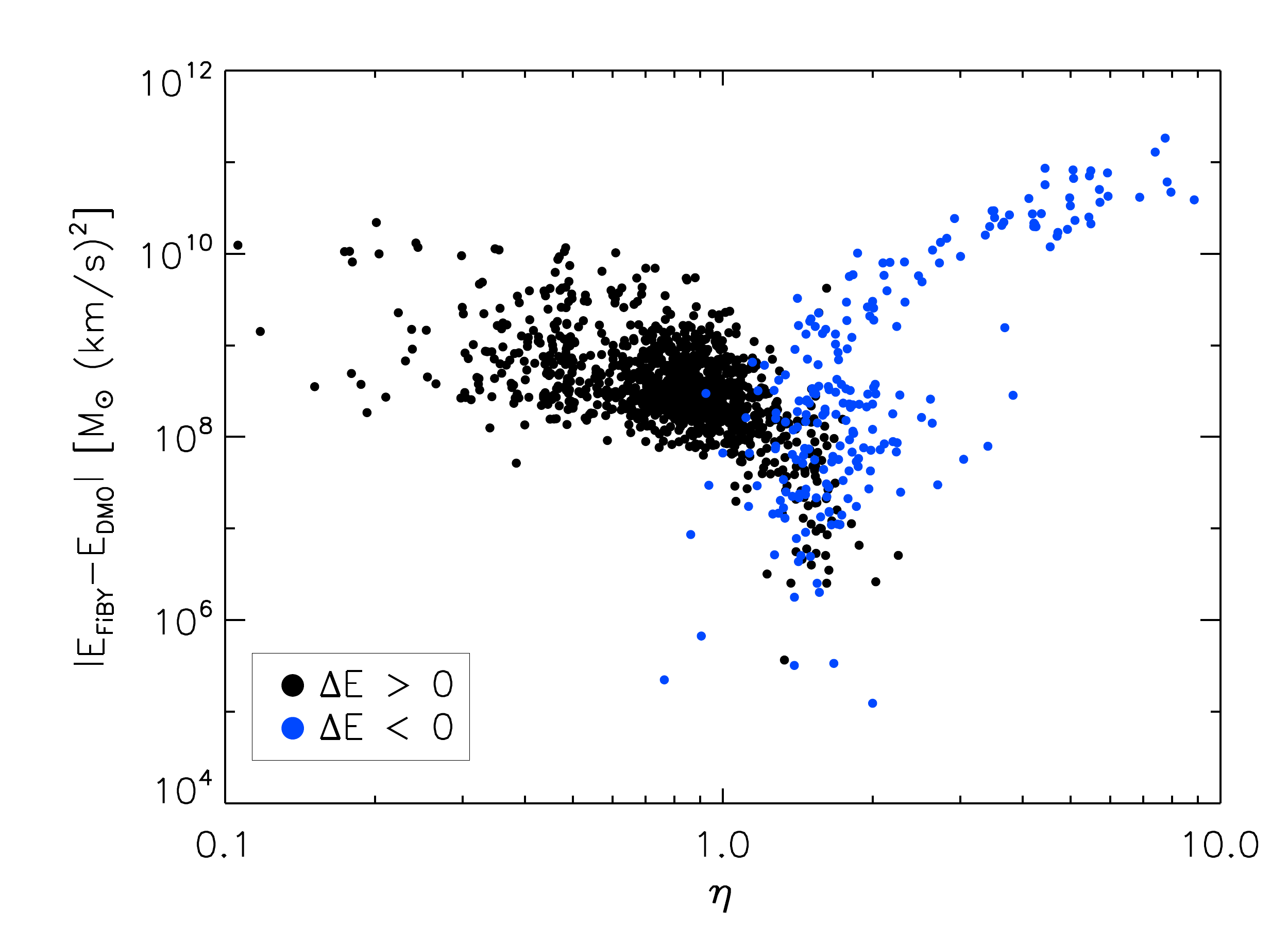}
\caption{Difference in total binding energy of the DM particles inside $\Rcut$ between the DMO and FiBY simulations as a function of $\eta$.  In black are haloes where the DMO halo is more bound, and in blue are the haloes where the FiBY halo is more bound.  We find that for $\eta < 0.7$ all FiBY haloes are less bound than their counterpart, and require between $10^8$ and $10^{10} \Msun (km/s)^2$ energy to be injected into the DM particles.}
\label{Fig:EBind}
\end{center}
\end{figure}

In studying the cycles of DM enhancement, it is important to quantify how much energy must be injected into the DM particles in order to explain the observed values of $\eta$.  This will provide insight into which physical models are capable of explaining the evolution of the DM profile.  We show in Figure \ref{Fig:EBind} the difference in total binding energy of the DM particles inside $\Rcut$ between the DMO and FiBY halo counterparts as a function of $\eta$.  The binding energy is calculated as the kinetic plus potential energy of each DM particle. Haloes where the DMO halo is more bound are plotted in black, while haloes where the FiBY halo is more bound are in blue.  We find that for all haloes with $\eta < 0.7$, the FiBY halo is less bound than it's DMO counterpart, with a difference in binding energy between $10^8$ and $10^{10} \Msun (km/s)^2$.  Using the parameters from our SN feedback model, this would require between $10^4$ and $10^6 \Msun$ in stars formed, or star formation rates over the past $5 \mathrm{\,Myr}$ between $0.002$ and $0.2 \Msun/\mathrm{yr}$, assuming the energy couples to the DM with perfect efficiency.  As this is obviously not the case, these should be taken as lower limits to the required star formation rates. We thus conclude that any mechanism to inject energy into the DM particles, such as supernova feedback, must transfer at least this much energy into the DM particles.

\begin{figure}
\begin{center}
\includegraphics[width = \columnwidth]{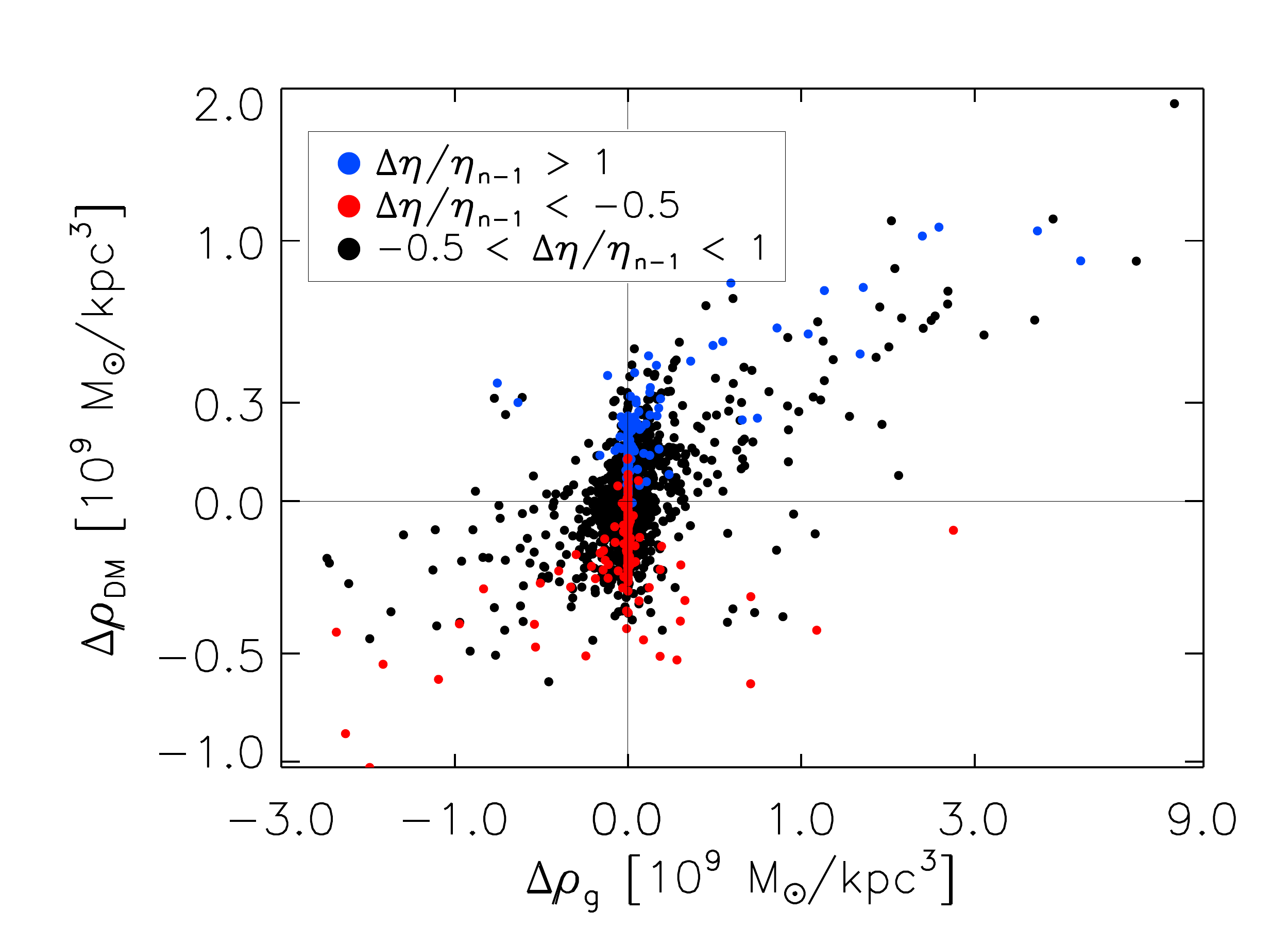}
\caption{Change in DM density versus the change in gas density between two snapshots, both measured within $\Rcut$. In general when a lot of gas moves in, the DM moves in as well, and vice versa. The points are coloured by the fractional change in $\eta$ between the snapshots.  }
\label{Fig:DRho}
\end{center}
\end{figure}

Finally, we wish to connect our simulation results to observable quantites: namely mass outflows and star formation rate.  In figure \ref{Fig:DRho}, we show the change in DM density between snapshots $n-1$ and $n$ as a function of the change in gas density, both measured as the average density inside $\Rcut$.  For infalling gas, where $\Delta \rho_g > 0$, the DM is nearly always moves in as well, and similarly when the gas moves out of the central region of the halo.  The haloes which show a large increase (decrease) in $\eta$ between the snapshots are coloured blue (red).  

\subsection{Adiabatic Contraction}
\label{Sec:MAC}
\begin{figure}
\begin{center}
\includegraphics[width = \columnwidth]{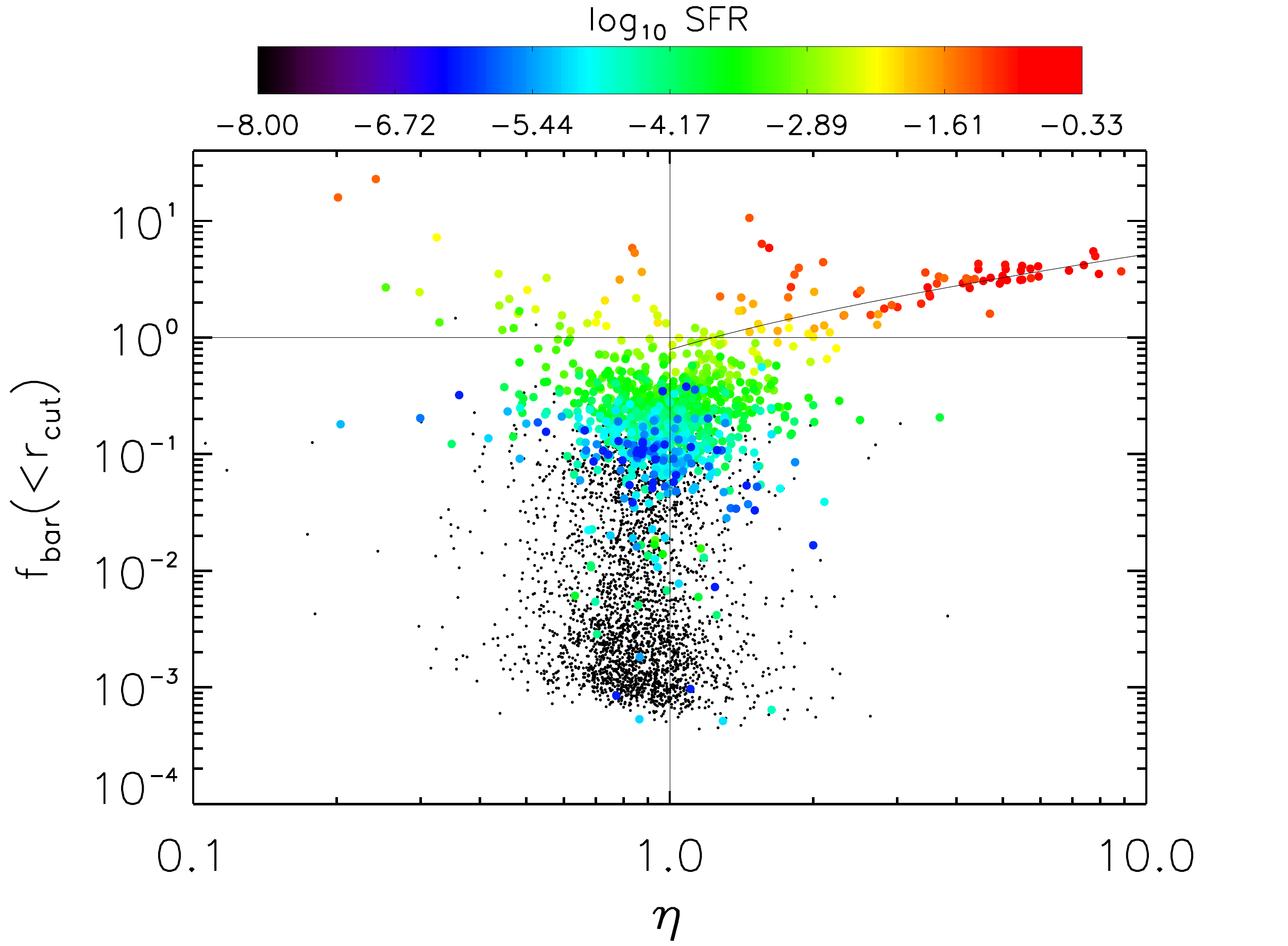}
\caption{Baryon ratio within $\Rcut$ versus $\eta$ for all haloes.  The points are coloured based on the instantaneous star formation rate inside $\Rcut$, with the small black points marking haloes with no current star formation.  Unsurprisingly, a large baryon ratio is required to have large values of $\eta$.  However, it is not a sufficient condition for enhancement, as is shown by the large number of haloes with $\eta < 1$ and $f_{\rm{bar}} > 1$.  These haloes have a lower star formation rate than their counterparts of the same baryonic fraction but large enhancements. The black curve is a power law fit of the form $\fb+1 = A\eta^B$ to the haloes with $\eta > 1 \mathrm{\,and\,} \fb > 1$, used in testing the adiabatic contraction model.}  
\label{Fig:BFrac}
\end{center}
\end{figure}

We have shown that haloes with large $\eta$ also have high SFR.  Hence, we expect there to also be a correlation between $\eta$ and the amount of baryons inside $\Rcut$ for haloes with $\eta > 1$.  We show in Figure \ref{Fig:BFrac} the correlation between the baryon ratio inside $\Rcut$ ($f_{\rm{b}} = M_{\rm{b}}(<r)/M_{\rm{dm}}(<r)$) and $\eta$.  We find that nearly all haloes with significant levels of enhancement have large baryon ratios in the central region of the halo -- implying that baryons are required to generate the enhancement.  However, the presence of baryons does not necessitate an enhancement in the DM profile, as is seen in the large subsample of haloes with baryonic fractions near unity but show a decrement in DM.  These haloes may be cases where the gas has just begun to fall back into a halo.  As the DM mass inside $\Rcut$ would be low for a halo with $\eta < 1$, a even small amount of gas would create a large baryonic ratio.  
	
Adiabatic contraction is a possible explanation for the enhancement in the dark matter in haloes with high central baryon fractions \citep[][]{Blumenthal86, Gnedin04, Abadi10}. We test whether the enhancement in seen in the dark matter density may be due to adiabatic contraction.   We discuss the adiabatic contraction model below, noting that it would also apply to adiabatic expansion if baryons slowly leave the halo.

The standard adiabatic contraction (SAC) model was first discussed by \citet{Blumenthal86}.  This model assumes spherical symmetry, circular orbits, homologous contraction, and that the specific angular momentum in a given mass shell is conserved.  Using the angular momentum as the conjugate momentum and the angular position as the coordinate, the adiabatic invariant is proportional to $r M(<r)$.  For baryons moving adiabatically in a halo, we can then write
\begin{multline}
\left[M^{\rm{DM}}_i(<r_i) + M^{\rm{b}}_i(<r_i) \right] r_i = \\ \left[M^{\rm{DM}}_f(<r_f) + M^{\rm{b}}_f(<r_f) \right] r_f,
\label{Equ:SAC}
\end{multline}
where the superscripts $DM$ and $b$ refer to the DM and baryonic mass inside the initial ($r_i$) and final radii ($r_f$).  This model, however, has been shown to systematically over-predict the mass profile in the inner region of simulated haloes \citep{Blumenthal86, Gnedin04}.  \citet{Gnedin04} relax the assumption of spherical orbits to allow for elliptical orbits in a modified adiabatic contraction (MAC) model.  This model uses the mass within an orbit-averaged radius, $\rt$, times the instantaneous radius $r$ as the conserved quantity instead of the specific angular momentum.  \citet{Gnedin04} propose a power-law parameterization between $\rt$ and $r$, such that 
\begin{equation}
\frac{\rt}{r_0} = A_0 \left( \frac{r}{r_0} \right)^w, 
\label{Equ:Rtilde}
\end{equation}
where the constants $A_0$ and $w$ were found by fitting to N-body simulations, and $r_0= 0.03 \rvir$.  With this relation, the adiabatic contraction model is given by 
\begin{multline}
\left[M^{\rm{DM}}_i(< {\rt}_i) + M^{\rm{b}}_i(< {\rt}_i) \right] r_i = \\ \left[M^{\rm{DM}}_f(< \rft) + M^{\rm{b}}_f(< \rft) \right] r_f.
\label{Equ:MAC}
\end{multline}
Using this modification, \citet{Gnedin04} report a better fit to the contraction of the DM in simulations with baryonic cooling than the SAC model.  

To test if our haloes follow the MAC model, we set the fitting parameters to $A_0 = 1.6$ and $w = 0.6$, though these have been shown to vary slightly from halo to halo \citep{Gustafsson06,Gnedin11} depending on the choice of baryonic feedback implementation, redshift, and halo mass.    \citet{Gnedin11} note that the scatter in $A_0$ and $w$ yields an accuracy of $\approx 10\%$ in predicting the dark matter profile.  Assuming the contraction is homologous with no shell crossing, we set 
\begin{equation}
M^{\rm{DM}}_f(<\rft) = M^{\rm{DM}}_i(<{\rt}_i).
\label{Equ:assume}
\end{equation} 
We use the DM mass profile from the DMO simulation at snapshot $n$ (scaled by $(\Omega_{\rm{M}}-\Omega_{\rm{b}})/\Omega_{\rm{M}}$) as the initial DM profile $M^{\rm{DM}}_i$, set $M^{\rm{b}}_i = 0$, and use the FiBY simulation at snapshot $n$ for the final DM and baryonic profiles.  We then solve for the initial radius, $r_i$, of the DM inside $r_f$.  

\begin{figure}
\begin{center}
\includegraphics[width = \columnwidth]{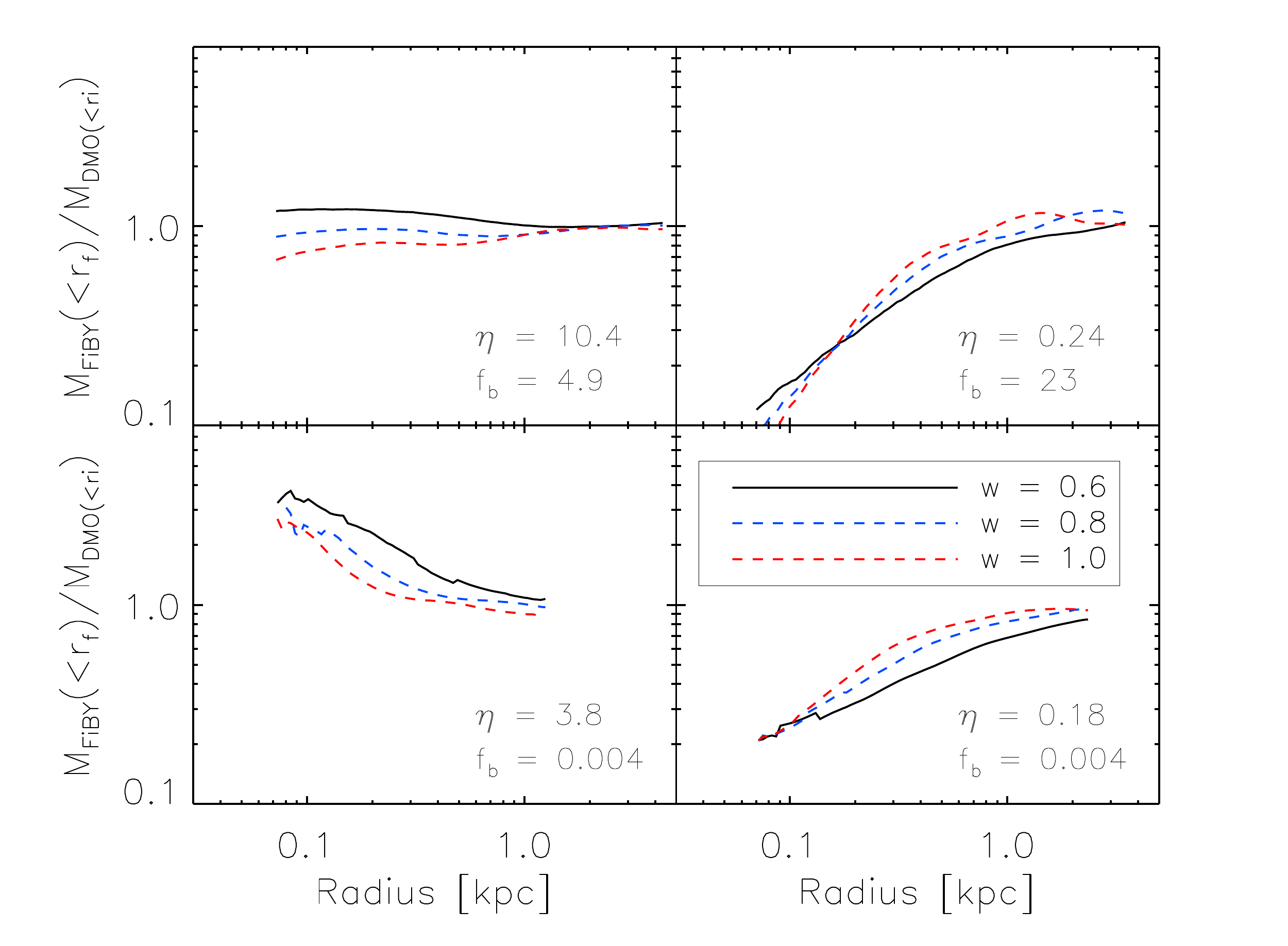}
\caption{Sample predicted mass ratios ($M^{\rm{DM}}_{\rm{FiBY}}(<\rft)/M^{\rm{DM}}_{\rm{DMO}}(<{\rt}_i)$) using the modified adiabatic contraction model for four haloes.  In the upper left, we show a halo with large $\eta$ and baryon ratio.  In the upper right, we show a halo with low $\eta$ but large baryon ratio.  In the lower panels, we show haloes with low baryon fraction and large (left) and small (right) $\eta$.  We vary the parameter $w$ in equation \ref{Equ:Rtilde} in the blue and red dashed curves, as suggested by \citet{Gnedin11}.  We find that only for haloes which have large baryon fractions and show $\eta > 1$ does the MAC model accurately describe the mass ratio.}
\label{Fig:MACradius}
\end{center}
\end{figure}

To test if the contraction observed in the FiBY simulation is due to adiabatic contraction, we test the validity of equation \ref{Equ:assume} using the DM profile from the FiBY run at snapshot $n$ as $M^{\rm{DM}}_f$, and the DM profile from the DMO run at snapshot $n$ as $M^{\rm{DM}}_i$.   We show in Figure \ref{Fig:MACradius} the ratio $M^{\rm{DM}}_{\rm{FiBY}}(<\rft)/M^{\rm{DM}}_{\rm{DMO}}(<{\rt}_i)$ as a function of $r_f$ for four haloes with different values of $\eta$ and baryon ratio inside $\Rcut$.  As can be seen, the choice of $w$ does not significantly affect the mass ratio profile.  

\begin{figure}
\begin{center}
\includegraphics[width = \columnwidth]{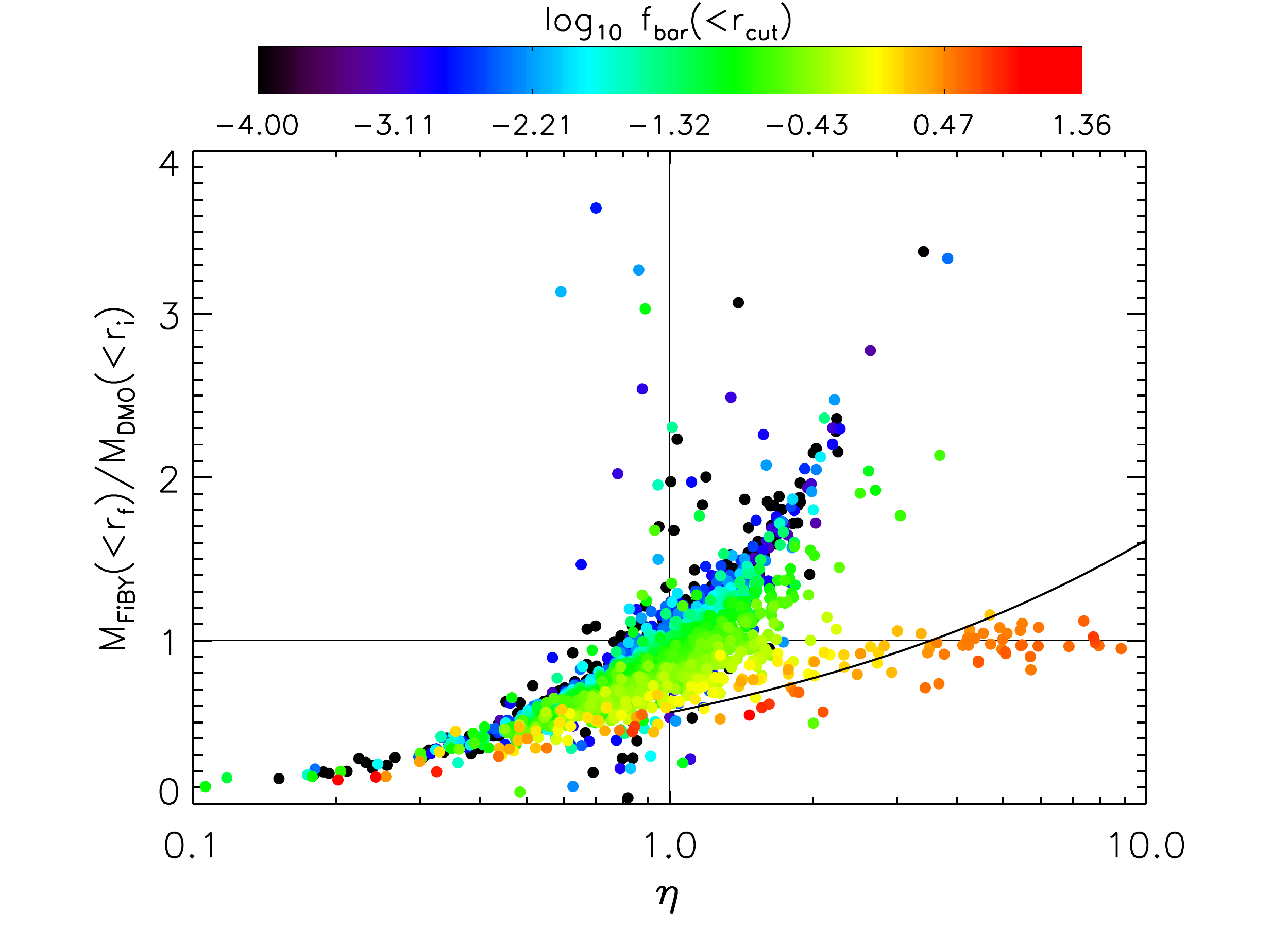}
\caption{Predicted mass ratio at $r_f = \Rcut$ using the modified adiabatic contraction model versus $\eta$.  We plot the ratio of the DM mass inside the orbit averaged final radius, $\rft$ from FiBY to the DM mass inside the orbit averaged initial radius, ${\rt}_i$ from the DMO run.  The initial radius is found using equation \ref{Equ:MAC}, and this ratio tests the assumption of equation \ref{Equ:assume}.  Each point is coloured by the baryon ratio inside $\Rcut$, and black points are for haloes with no baryons inside $\Rcut$.  For haloes with $\eta > 2$ and $f_{\rm{bar}} \gtrsim 0.63$, the model works to within $\approx 20\%$.  However, many haloes with $\eta \approx 2$ show low baryon fractions and more dark matter mass inside $\rft$ than the MAC model predicts.  The solid black curve denotes the prediction of the SAC model, assuming a power law relation between $\fb+1$ and $\eta$ for haloes with $\eta > 2$ and $\fb > 1$.}
\label{Fig:ModAC}
\end{center}
\end{figure}

In Figure \ref{Fig:ModAC} the same ratio evaluated at $r_f = \Rcut$ is plotted for each halo as a function of $\eta$.  The points are coloured by the baryon ratio inside $\Rcut$.  We find that only for haloes with $\eta > 1$ and baryon ratio inside $\Rcut$ greater than $\approx 0.6$ does the MAC model work well.  Without enough baryons, it is impossible to effectively modify the DM profile to create a strong enhancement.  

It is instructive also to take the observed relation between the baryon ratio, $\fb$, seen for haloes with $\eta > 2$ and $\fb > 1$ in Figure \ref{Fig:BFrac} (black curve). If we fit the data with a power law such that $\fb+1 = A\eta^B$, one can derive (see Appendix A) another power law relating the mass ratio plotted in Figure \ref{Fig:ModAC} to $\eta$: 
\begin{equation}
\frac{M^{\rm{FiBY}}(<r_f)}{M^{\rm{DMO}}(<r_i)} = \frac{1}{A} \eta^{1-B}.
\label{Equ:MACFit}
\end{equation}
Using the best fit values of $A = 1.786$ and $B = 0.540$,  we show this curve as the solid black curve in figure \ref{Fig:ModAC}, and find that it does a decent job of predicting where haloes with $\eta > 2$ and $\fb > 1$ lie in this plane, confirming that the enhancement in these haloes is due to adiabatic contraction.  We note though, that it slightly over-predicts the mass ratio at the largest values of $\eta$.  This is unsurprising, as the simple derivation here assumed the SAC model, which is known to over-predict the enclosed DM mass at high baryon ratios.

\begin{figure}
\begin{center}
\includegraphics[width=\columnwidth]{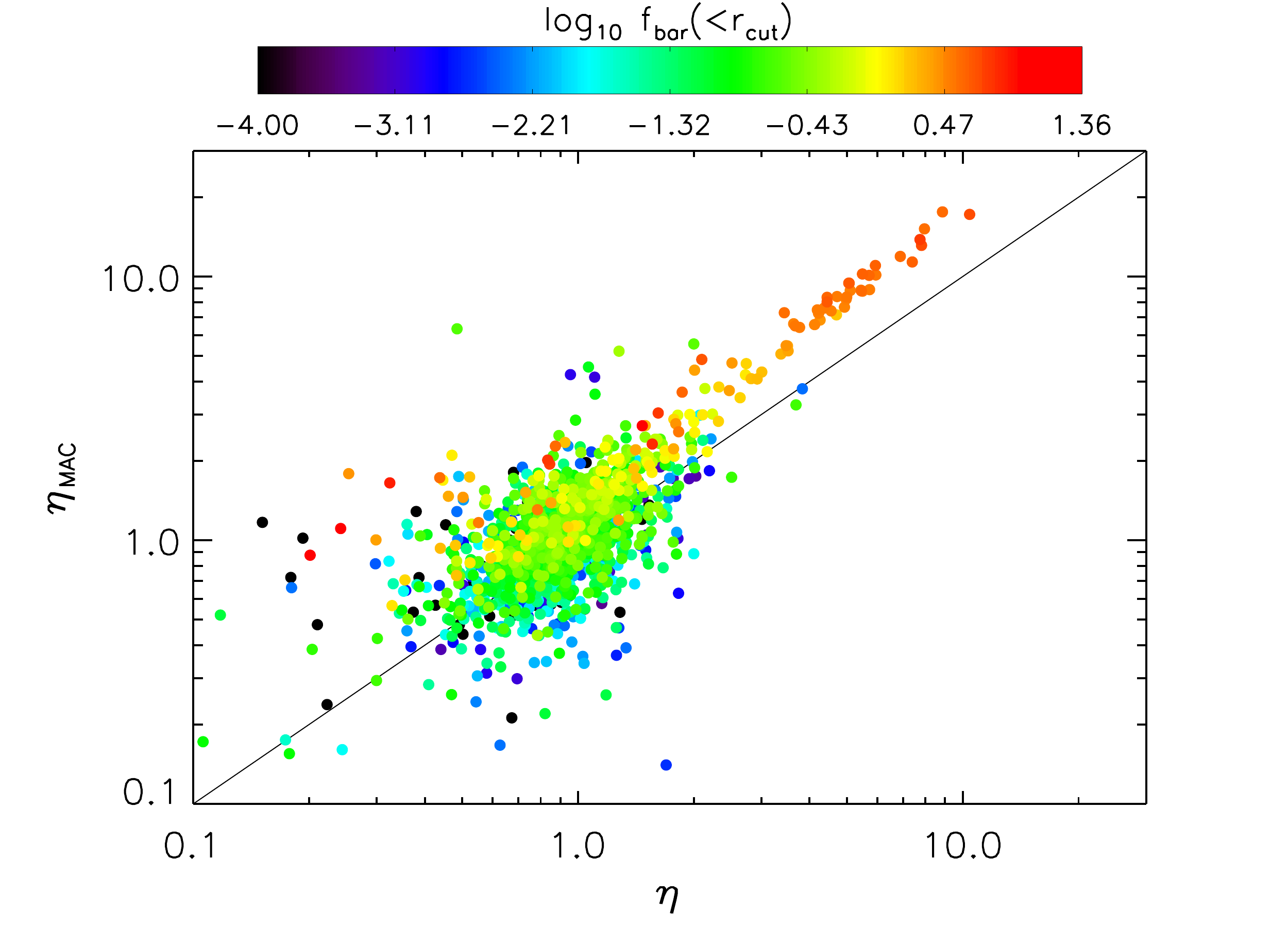}
\caption{Predicted $\eta_{\rm{MAC}}$ versus $\eta$ using the MAC model of equation \ref{Equ:MACEtap} where the baryon and DM mass profiles from the FiBY simulation are used as the initial inputs to predict the FiBY DM mass profile in the next snapshot.  For the haloes in which the model is expected to do well -- haloes with a high baryon ratio -- the model matches the expected slope, but overpredicts $\eta_{\rm{MAC}}$ by $\approx 10\%$. }
\label{Fig:MACEtap}
\end{center}
\end{figure}

It is also instructive to see how the MAC model may predict the evolution of the dark matter profile in the FiBY simulation from snapshot $n-1$ to snapshot $n$.  Previously, we assumed that all the baryons are added directly to the DMO mass profile by setting $M^{\rm{b}}_i = 0$ and using the DMO simulation for $M^{\rm{DM}}_i$ at snapshot $n$.   However, the DM profile of the FiBY simulation at snapshot $n-1$ has already been affected by the baryons present at that time, and so is not necessarily comparable to the DMO mass profile at snapshot $n$.  Therefore, we apply equation \ref{Equ:MAC} to the evolution of a FiBY halo between snapshots $n-1$ and $n$.  We use the baryon and dark matter mass profiles at snapshot $n-1$ in the FiBY simulation as our initial $M^{\rm{DM}}_i(< {\rt}_i)$ and $M^{\rm{b}}_i(< {\rt}_i)$.   Using equation \ref{Equ:assume} and the baryon mass profile at snapshot $n$, we then solve for $r_i$ and test our assumption using the ratio 
\begin{equation}
y(r) = M^{\rm{DM}}_{\rm{FiBY}}(<\tilde{r})/M^{\rm{DM}}_{\rm{FiBY}}(<{\rt}_i).
\end{equation}
With this definition, $y(r)$ captures how well the MAC model is able to predict the time evolution of our FiBY haloes.
Evaluating this ratio at $r = \Rcut$, we can make a prediction for the enhancement, 
\begin{equation}
\eta_{\rm{MAC}} = \eta / y(\Rcut).
\label{Equ:MACEtap}
\end{equation}
 We compare the predicted value for $\eta{\rm{MAC}}$ to the measured value in Figure \ref{Fig:MACEtap}.  We find that for high values of $\eta$, the model matches the correct slope, but with a systematic offset of about $10\%$ too high.  The offset may be from our choices of $A_0$ and $w$, though we note that our offset is well within the expected error discussed in \citet{Gnedin11}.  The offset may also imply that while the adiabatic contraction is dominating the evolution of $\eta$ in strongly enhanced haloes, other factors may be inhibiting enhancement of DM; we explore some of these possibilities in Section \ref{Sec:BarOut}.

\subsection{Removal of Baryons}
\label{Sec:BarOut}
\subsubsection{Reionization}
\begin{figure}
\begin{center}
\includegraphics[scale = 0.375]{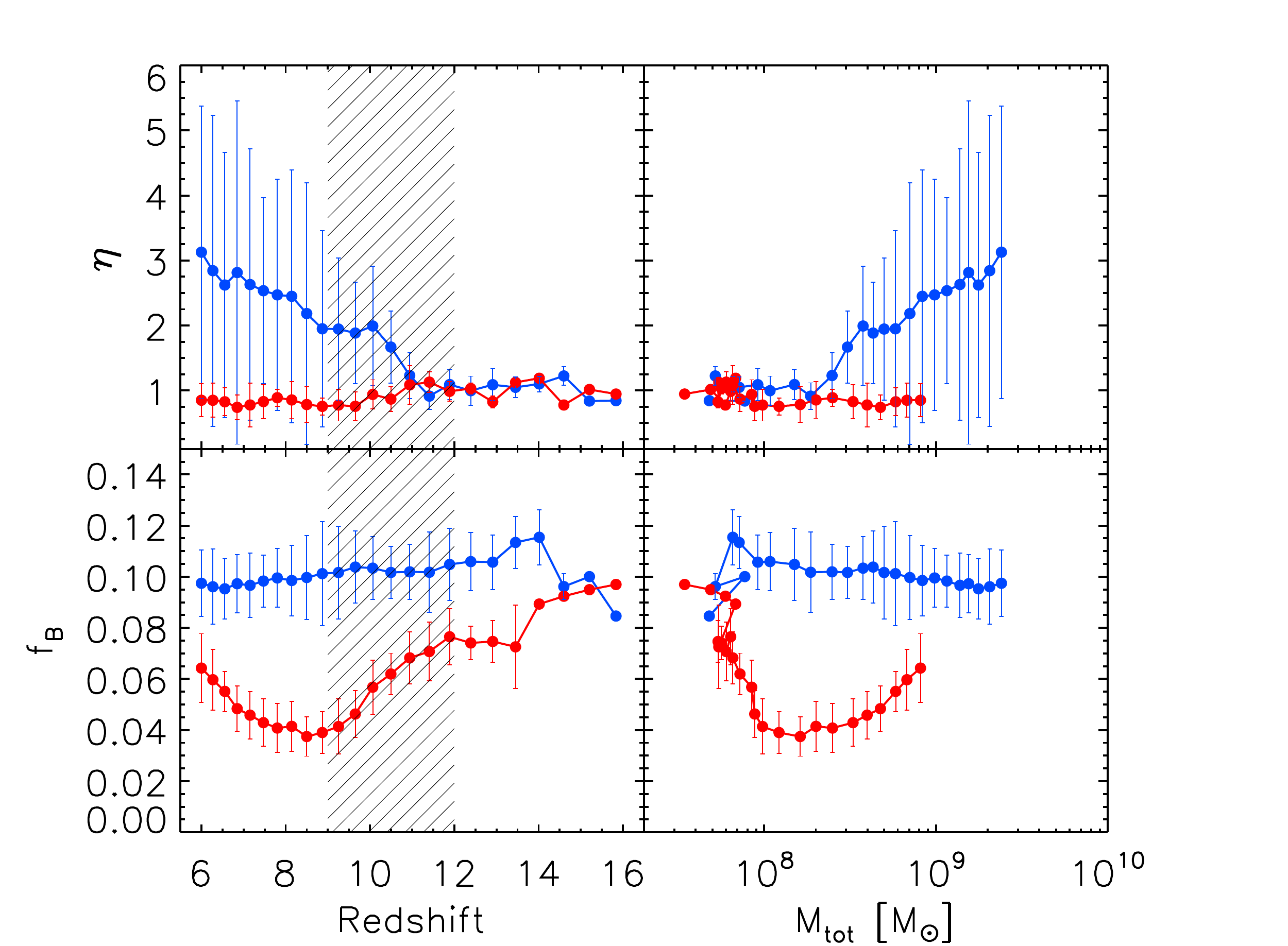}
\caption{Evolution of the mean $\eta$ with redshift (top left) and halo mass (top right) for haloes in two mass bins.  In blue are haloes with $\Mtot > 1.5\times10^9\Msun$ at $z=6$ and in red are haloes with $6.5\times10^8\Msun < M < 1.5\times10^9\Msun$ at $z=6$.  The bottom left panel shows the evolution of the mean baryon fraction inside $\Rvir$ for the same two samples.  The gray hashed region marks the epoch of reionization, during which baryons evaporated out of the lower mass haloes, prohibiting the buildup of a cuspy DM profile and a large value of $\eta$. The error bars show the standard deviation in each halo sample.}
\label{Fig:Redshift}
\end{center}
\end{figure}

As is seen in Figure \ref{Fig:EtaHist}, not every halo of a given mass shows a similar enhancement (or decrement) of dark matter.  We examine this further in Figure \ref{Fig:Redshift}, where we show in the upper left panel the redshift evolution of the mean $\eta$ for two halo samples: those which at $z=6$ have $\Mtot > 1.5\times10^9\Msun$ (blue points) and those with $6.5\times10^8\Msun < \Mtot < 1.5\times10^9\Msun$ (red points) -- roughly a factor of 2 difference in halo mass at $z=6$.   To obtain the redshift evolution, we follow the progenitors of the haloes in these two samples, and plot the mean $\eta$ for each sample at each redshift.  We also show the evolution of the mean $\eta$ as a function of the mean total mass in the upper right panel.  We find that when the haloes are below $\Mtot \approx 3 \times 10^8 \Msun$, they have similar $\left<\eta\right>$ and $\sigma_\eta$.  These low mass haloes have on average $\eta$ slightly less than unity.  The tracks diverge at $\Mtot \approx 3\times10^8 \Msun$; the higher mass sample develops a stronger enhancement on average as well as more halo to halo variation.  As the haloes in this mass bin reach $\Mtot = 3 \times 10^8 \Msun$ earlier than the lower mass bin haloes, the halo formation time likely plays a role in the structure of the central dark matter profile.

From the redshift evolution, we find that the high mass sample diverges from the lower mass sample at $z\approx 11$: during the epoch of reionization.  The bottom left panel of Figure \ref{Fig:Redshift} shows the mean baryon fraction inside $\Rvir$ in the haloes as a function of redshift.  The shaded region denotes the time when our reionization model is active. The haloes in the lower mass bin lose baryons during reionization while the higher mass haloes do not.  Without enough baryons to cool and collapse to the centre, the lower mass haloes are unable to generate an enhancement of $\eta$.  

\begin{figure}
\begin{center}
\includegraphics[width = \columnwidth]{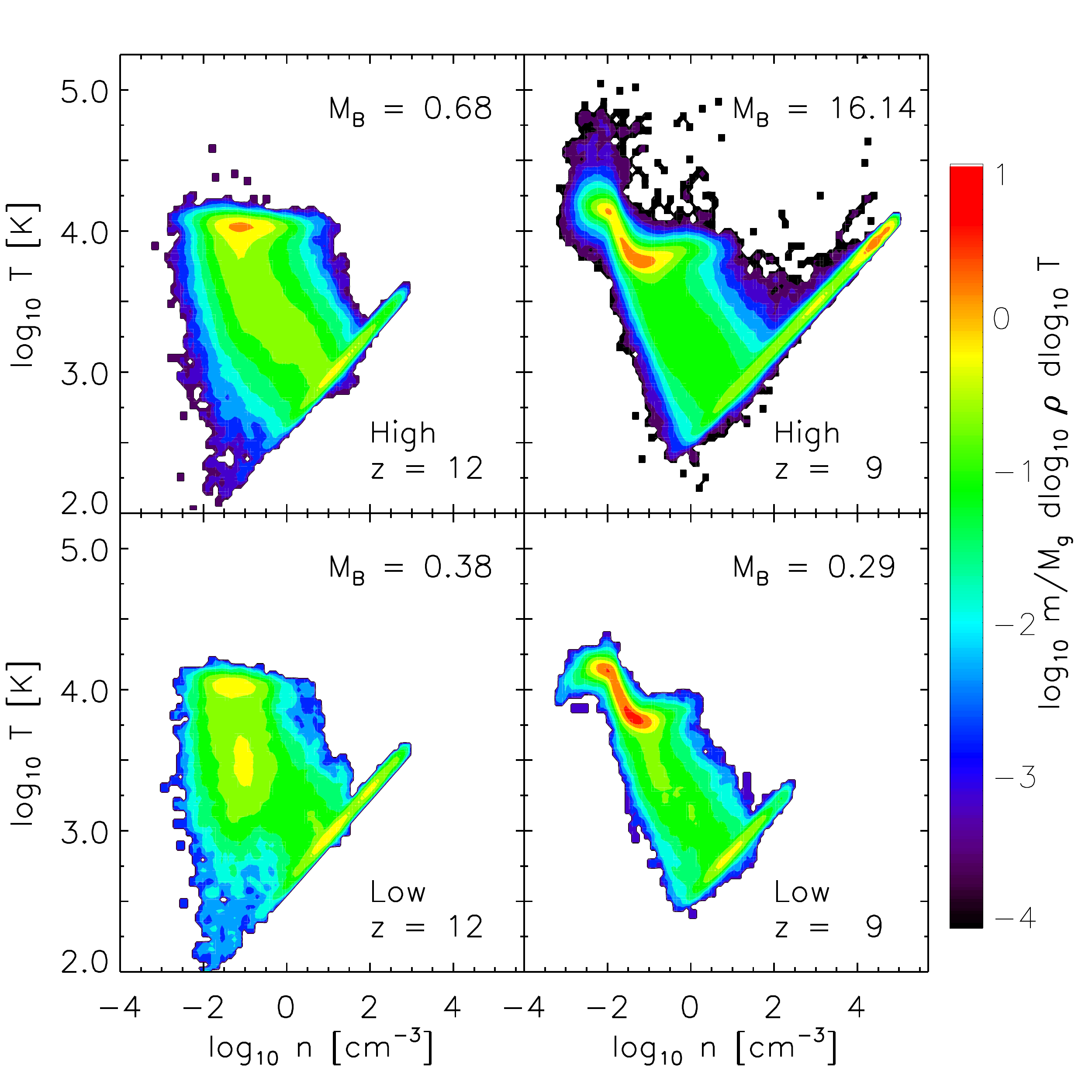}
\caption{Phase diagram for the gas in haloes in two mass bins before and after reionziation.  The mass bins are the same as those in Figure \ref{Fig:Redshift} and are based on the total mass of the halo at $z=6$.  The baryonic mass inside $\Rcut$ is listed in the upper right of each panel in units of $10^6 \Msun$.  We find that haloes in the higher mass bin (top row) retain dense gas, while the haloes in the lower mass bin do not (bottom row).  Hence, the high mass haloes have on average larger values of $\eta$ than the lower mass haloes.}
\label{Fig:PhaseReion}
\end{center}
\end{figure}

The effect is also seen in Figure \ref{Fig:PhaseReion} which shows phase diagrams of the gas before and after reionization.  The top (bottom) row shows the $\rho - T$ diagram for all the gas inside $\Rvir$ in the higher (lower) mass bins.  The mean baryonic mass inside $\Rcut$ is listed in each panel, and shows that the higher mass haloes increase in baryonic mass during reionization while the lower mass haloes do not.  These more massive haloes contained enough high-density gas to self-shield against the UV background during reionization allowing further baryonic collapse and modifications to the DM profile.  The lower mass haloes do not have as much high density gas, and so more gas is heated by the UV background and unable to collapse.  Hence, in these haloes we do not see the buildup of strong DM enhancements.  

\subsubsection{Supernovae Outflows}
\label{Sec:Outflow}
As shown in \citet{Pontzen12} and \citet{Ogiya12}, rapid outflows driven by stellar feedback can cause rapid changes in the halo potential.  As the potential becomes shallower, the dark matter gains kinetic energy, thereby destroying a central cusp.  \citet{Pontzen12} show that for rapid changes in the potential, even if the potential returns to its initial state the DM particles will still have a net gain in total energy and not return to their initial configuration.  

To test the effectiveness of outflows in removing dark matter from the center of our haloes, we implement the model described in \citet{Pontzen12}.  For each halo at snapshot $n$, we initialize tracer particles with the location and velocity of the dark matter particles at the previous snapshot, $n-1$.  These tracer particles are then placed in a spherical potential, $V(r)$ calculated from the spherically averaged total density profile at snapshot $n-1$.  We then assume the potential instantaneously changes to the potential at snapshot $n$.  The new total energy per unit mass of a tracer particle at time $n$ is given as 
\begin{equation}
E_n = \frac{1}{2} v_{n-1}^2 + V_n(r_{n-1}).
\end{equation}
We assume that the particle's specific angular momentum is conserved between snapshots, so that $j_{n} = j_{n-1} = \vec{r}_{n-1} \times \vec{v}_{n-1}$.  Given the new total energy of a tracer particle and its angular momentum and defining an effective potential $V_{\mathrm{eff}}(r) = V(r) + 0.5 (j/r)^2$, we can integrate over the new particle orbit to obtain the radial probability distribution, $p(r)$:
\begin{equation}
p(r)  =  
\begin{cases}
\frac{1} {(E-V_{\mathrm{eff}}(r))^{1/2} \int \!  (E - V_{\mathrm{eff}}(r))^{-1/2}\,\mathrm{d}r} & E-V_{\mathrm{eff}}(r) > 0 \\
0 & E-V_{\mathrm{eff}}(r) \leq 0.
\end{cases}
\end{equation}
By summing the probability distribution over all tracer particles, we can predict the final density profile at snapshot $n$ for the tracer particles as
\begin{equation}
\rho_{\rm{model}}(r) =  \frac{m_{\rm{DM}}}{4 \pi r^2} \sum_l p_l(r).
\label{Equ:Pontzen}
\end{equation}
With the predicted density profile, we can then integrate the profile to find the DM mass inside $100 \pc$ and use this to predict $\eta$ at snapshot $n$.   

\begin{figure}
\begin{center}
\includegraphics[width=\columnwidth]{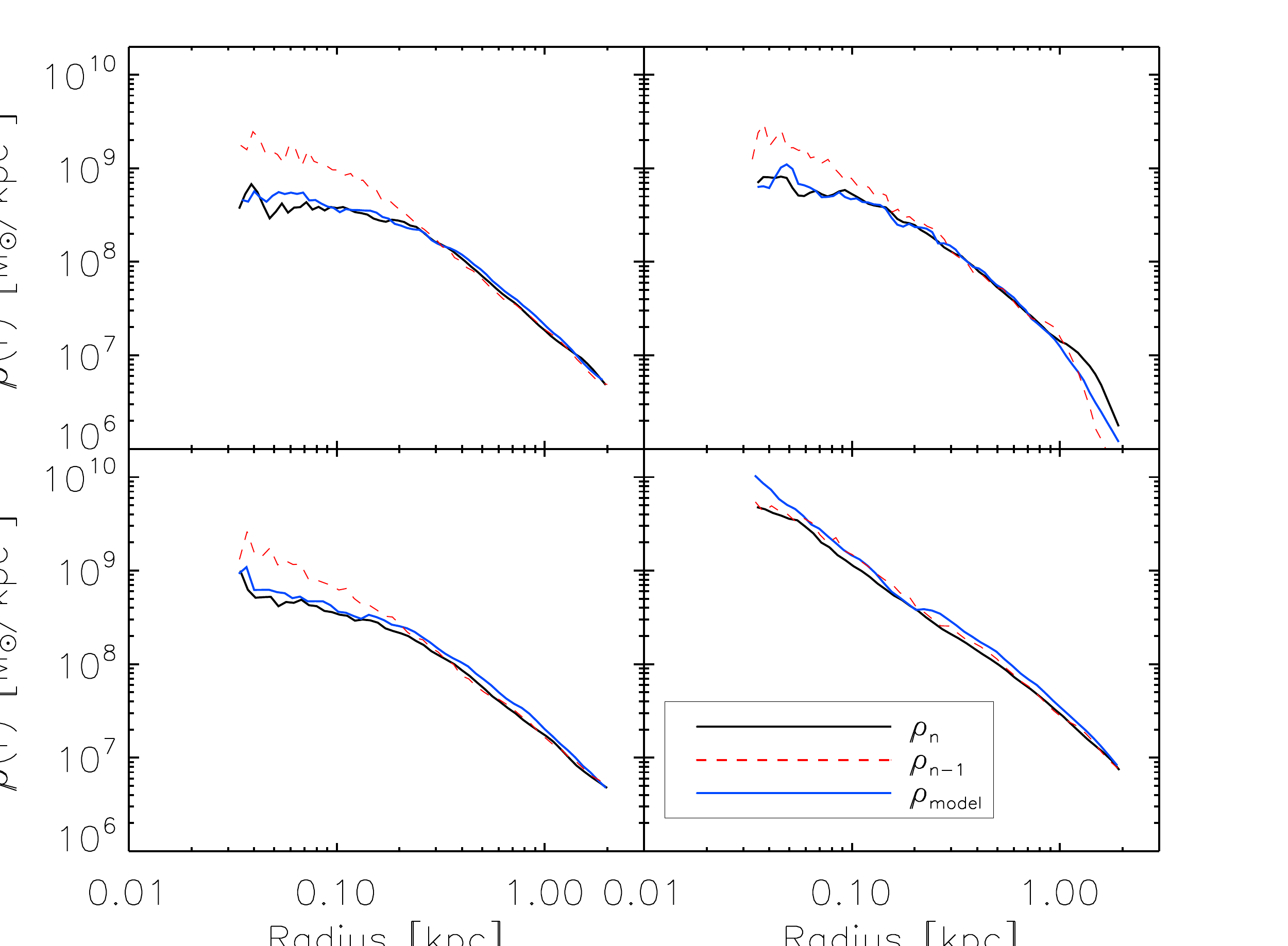}
\caption{DM density profiles for four haloes which have a net loss of baryons and a decrease in $\eta$ between snapshots $n-1$ (red dashed) and $n$ (black solid). In blue is shown the density profile predicted by equation \ref{Equ:Pontzen}. We find that the outflow model does a good job at predicting the response of the DM to the baryonic outflow.}
\label{Fig:Pontzen}
\end{center}
\end{figure}

We show in Figure \ref{Fig:Pontzen} the predicted density profile of Equation \ref{Equ:Pontzen} for four haloes which have a net outflow of baryons and a decrease in $\eta$ by about $60\%$ between two snapshots.  Three of them do extremely well in reproducing the observed DM profile at snapshot $n$, and can predict $\eta$ to within a factor of $\approx 50\%$.  However, the bottom right panel shows a halo where the decrease in $\eta$ is driven by a factor of $2$ growth in the DMO halo and not by a strong change in the FiBY halo.  In this case, the outflow model still matches the FiBY density profile well, even though there is little change in the density profile.  Thus, we conclude that when baryons are rapidly leaving the halo, the outflow model does a good job of predicting the resulting DM density profile.

\subsubsection{Dynamical Friction Heating}
\label{Sec:DF}
Another proposed method of destroying a central DM enhancement is via dynamical friction from infalling galaxies \citep{ElZant01, ElZant04, Tonini06, RomanoDiaz08, Maccio12}.  \citet{Lackner10} proposed a model for dissipationless energy transfer from infalling stars to a host galaxy's dark matter.  In this scenario, the orbital energy lost by the stars is deposited into a shell of DM, causing it to expand and the density in the shell to decrease.  We use this model to explore the possibility that dynamical friction is the dominant source of DM expansion instead of the outflow model of the previous section.  

\begin{figure}
\begin{center}
\includegraphics[width = \columnwidth]{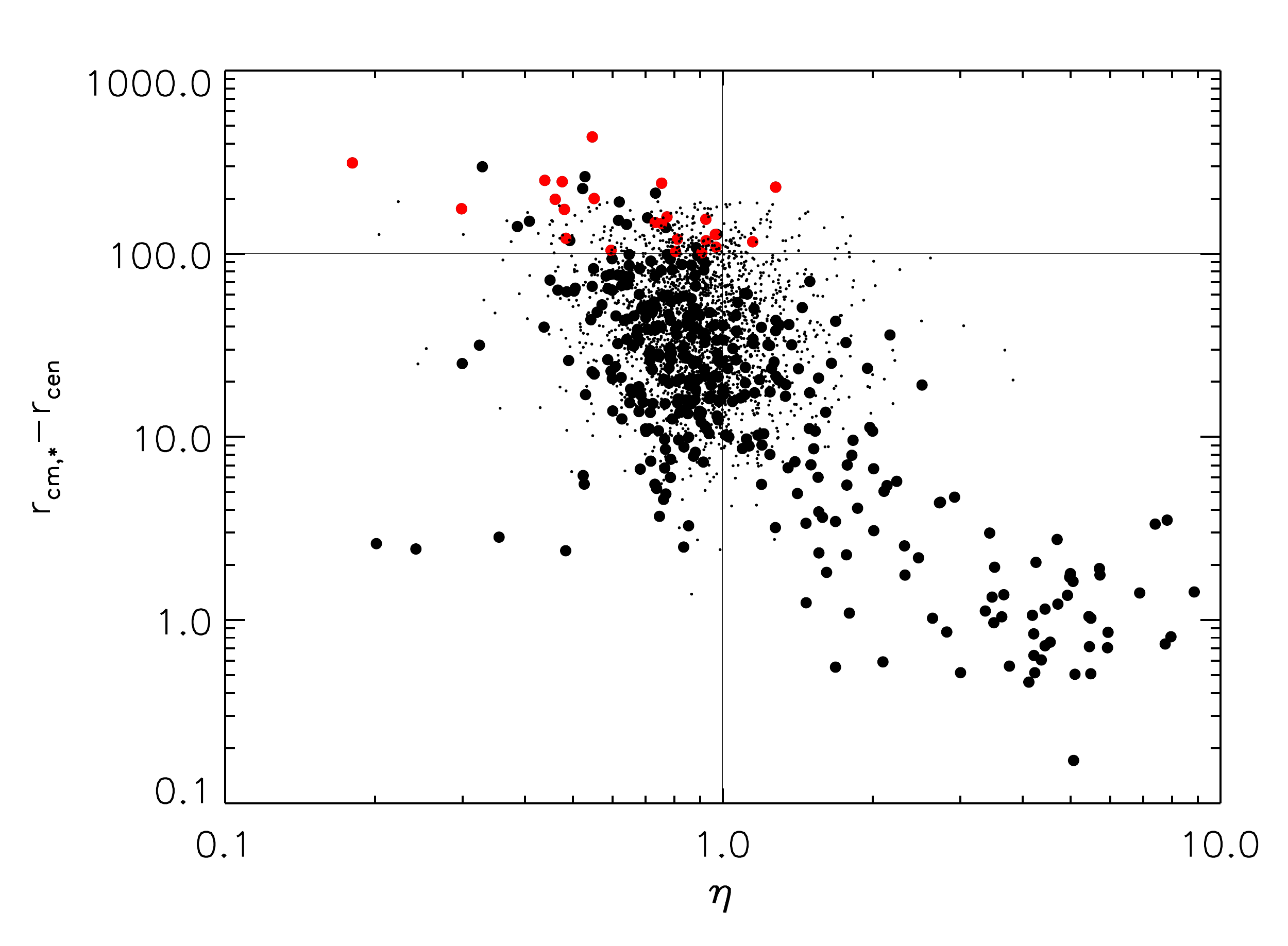}
\caption{Offset between the centre of mass of the stars and the halo centre (in physical parsecs) versus halo enhancement for all haloes at all redshifts in our sample.  The large points are for haloes with at least $50$ star particles inside $\Rcut$, and the small points have fewer than $50$ star particles.  We find that haloes with large enhancements generally have small stellar offsets.  Note that the gravitational smoothing parameter reaches a maximum of $33 \rm{\,pc}$ (physical) at $z=6$, and so many offsets are less than one smoothing length.   The red points mark haloes with at least $50$ star particles and which have an offset larger than $100\rm{\,pc}$ currently but have an offset less than this in the next time step.  They are the candidates to which we apply the dynamical friction model. }
\label{Fig:DMOffset}
\end{center}
\end{figure}

In Figure \ref{Fig:DMOffset} we show the offset between the centre of mass of the stars and the centre of the halo (i.e. the centre of the global potential well) for all haloes.  We find a general trend to decreasing offset with increasing $\eta$.  We also identify haloes (red points) where the offset is larger than $\Rcut$ in one snapshot $n-1$ and smaller than $\Rcut$ in the next snapshot $n$.  These $23$ haloes are then studied further to see if dynamical friction from the infalling stars may affect the evolution of $\eta$.

Following Equation 11 in \citet{Lackner10} we define the difference between the total energy of the DM and stellar shells as
\begin{eqnarray}
\Delta E & = & \frac{G}{12 \Rcut} \Bigl\{ (\Delta R^{\prime}-\Delta R) \Bigl[ 2 {\Mdm}^{2} +   \nonumber \\
 & & 3 M_{\rm{int}} \Mdm + M_* \Mdm + \frac{\Rcut \Mdm}{\Delta R^{\prime}} M_{\rm{ext}} \Bigr] \nonumber \\
 & & -3 \Delta R M_* (\Mdm + 2 M_{\rm{int}} + M_*) \Bigr\}
 \label{Eq:Lackner}
 \end{eqnarray}
where $\Delta R$ is the initial DM shell width which we set as $20 \pc$ centered on $\Rcut$, $\Mdm$ is the DM mass in the shell, $M_*$ is the stellar mass crossing $\Rcut$, $M_{\rm{int}}$ and $M_{\rm{ext}}$ are the total mass interior and exterior to the shell.  Equipartition of energy between the DM shell and the stars allows us to set $\Delta E = 0$; we then solve iteratively for the new shell width, $\Delta R^\prime$, giving a new density in the shell.  From that new density, we then calculate how much DM is still inside $\Rcut$ to obtain $\eta_{\rm{DF}}$ by dividing the mass inside $\Rcut$ predicted by Equation \ref{Eq:Lackner} by the mass in the DMO counterpart halo at snapshot $n$.  We note that in our calculations, we have assumed all the stars start outside $\Rcut$ if the stellar center of mass is outside $\Rcut$. Thus our estimate of the mass lost due to dynamical friction is an upper limit.

\begin{figure}
\begin{center}
\includegraphics[width = \columnwidth]{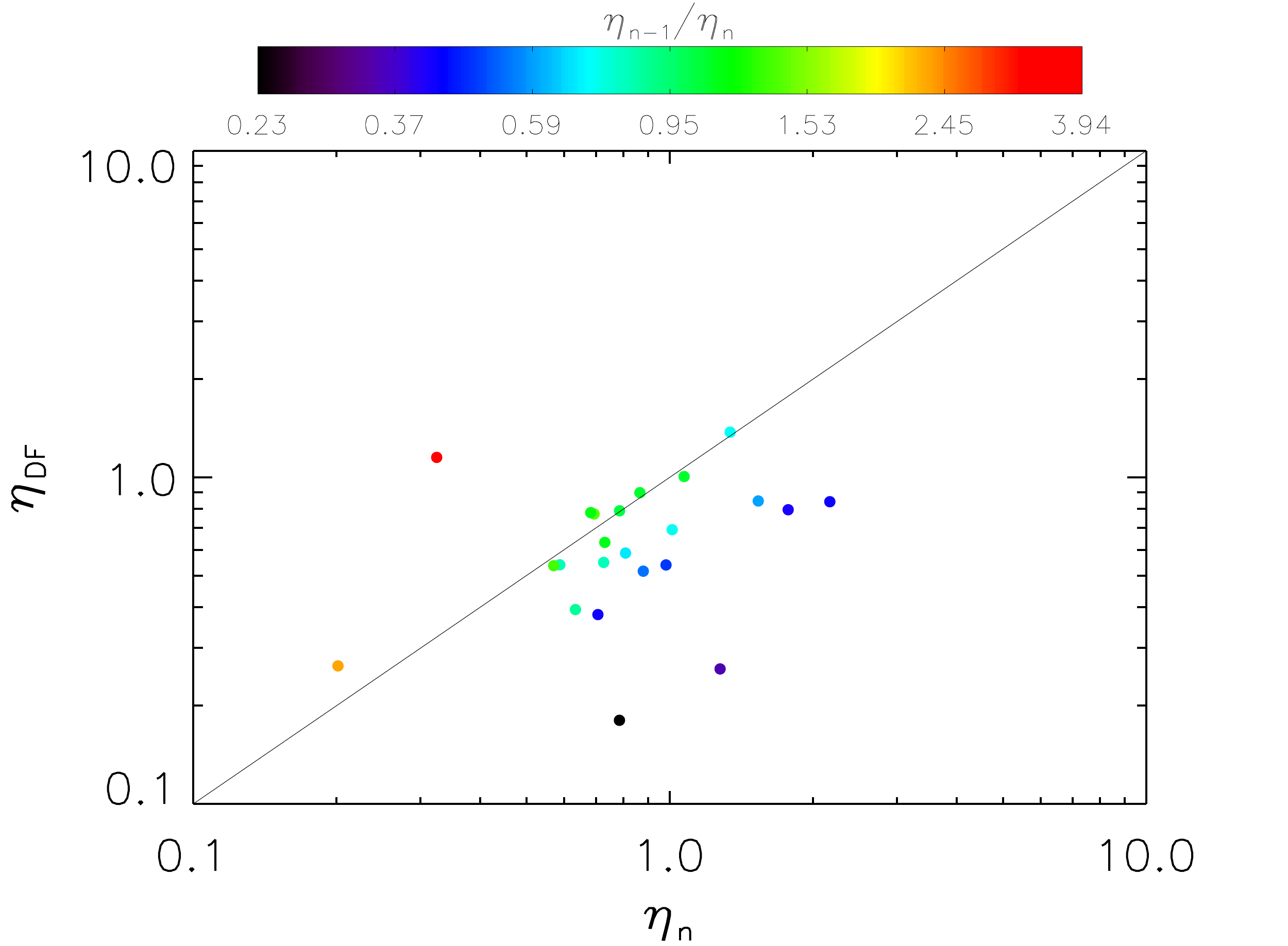}
\caption{Predicted $\eta_{\rm{DF}}$ from the dynamical friction model using equation \ref{Eq:Lackner} versus actual $\eta_n$ in the following snapshot for the same halo sample.  The model is applied to haloes which at snapshot $n-1$ have a stellar offset larger than $\Rcut$ and a stellar offset less than $\Rcut$ at snapshot $n$, as well as having at least $50$ star particles (the red points in Figure \ref{Fig:DMOffset}).  The points are coloured by the ratio of enhancements between snapshot $n-1$ and $n$: $\eta_{n-1}/\eta_n$.  The model accurately predicts $\eta_n$ only for haloes which show little change in enhancement between snapshots.}
\label{Fig:OffsetNext}
\end{center}
\end{figure}

In Figure \ref{Fig:OffsetNext} we show the enhancement, $\eta_{\rm{DF}}$, predicted by equation \ref{Eq:Lackner} versus the measured $\eta_n$.   The points are coloured by the ratio $\eta_{n-1}/\eta_n$, showing which haloes have actually undergone a change in enhancement between the two snapshots.  The first point to note is that most haloes have $\eta_{n-1}/\eta_n < 1$, meaning the enhancement increases with time.  We would expect these haloes to not be well fit by the dynamical friction model, and indeed, they are not; $\eta_{\rm{DF}}$ is less than $\eta_n$ by factors of a few for these haloes.  We find that the mass lost due to dynamical friction is very small -- of order $0.05\%$ of the DM mass inside $\Rcut$ -- and so this model predicts very small change to the mass inside $\Rcut$ between snapshots.  Hence, we find that $\eta_{\rm{DF}} \approx \eta_n$ only for haloes which have very little change in $\eta$ between snapshots $n-1$ and $n$.

This result can be explained by straightforward energy arguments.  Using a toy NFW profile with $\Mtot = 10^9 \Msun$, we can add a cusp to the density profile inside $\Rcut$ having a logarithmic slope of $-2$.  If the mass in the cusp is only $3\%$ of the total dark matter mass we can then calculate the energy required to remove this mass from the halo.  Then, assuming $10\%$ of the baryons are in stars which are moved in from infinity to $50 \pc$, we can calculate the maximum amount of energy available for unbinding the DM cusp.  Using these parameters, we find that the energy required to unbind a cusp leaving only an NFW profile behind is $8.7$ times as much as the potential energy released by the infalling stellar clump. 

\subsection{Model Prediction}
To test the predictive power of the three physical models described in Sections \ref{Sec:MAC} and \ref{Sec:BarOut}, we use the models to predict the enhancement, $\eta_p$, at snapshot $n$ given the state of the halo at snapshot $n-1$ and compare it to the measured $\eta$ from the simulation.  We track the movement of the gas between the two snapshots to determine which models we apply to each halo.

For haloes which show a net inflow of gas (within $\Rcut$) from one snapshot to the next, we used the MAC model described in section \ref{Sec:MAC}.  We use the properties of the FiBY halo at $n-1$ as the initial DM and baryonic profiles, and the baryon profile at $n$ as the final baryon profile.   We use equation \ref{Equ:MAC} to solve for $r_i$, and then, assuming equation \ref{Equ:assume}, we predict the DM density profile, $\rho_p(r)$ at $n$.  For haloes which show a net loss of gas (again within $\Rcut$), we apply the rapid outflow model of section \ref{Sec:Outflow} to create $\rho_p(r)$ at $n$ using Equation \ref{Equ:Pontzen}.  

Finally, for all haloes, we track how many stars inside of $\Rcut$ at $n$ were outside of $\Rcut$ at $n-1$ (either in the same halo or one that merged with the halo between $n-1$ and $n$).  Using the total mass of these stars and $\rho_p$, we can then apply equation \ref{Eq:Lackner} to obtain the final predicted DM density profile.  With this profile, we can calculate the predicted DM mass inside $\Rcut$ at $t_n$ to obtain $\eta_p$ by dividing by the DM mass inside $\Rcut$ from the DMO simulation at $t_n$. 

\begin{figure}
\begin{center}
\includegraphics[width=\columnwidth]{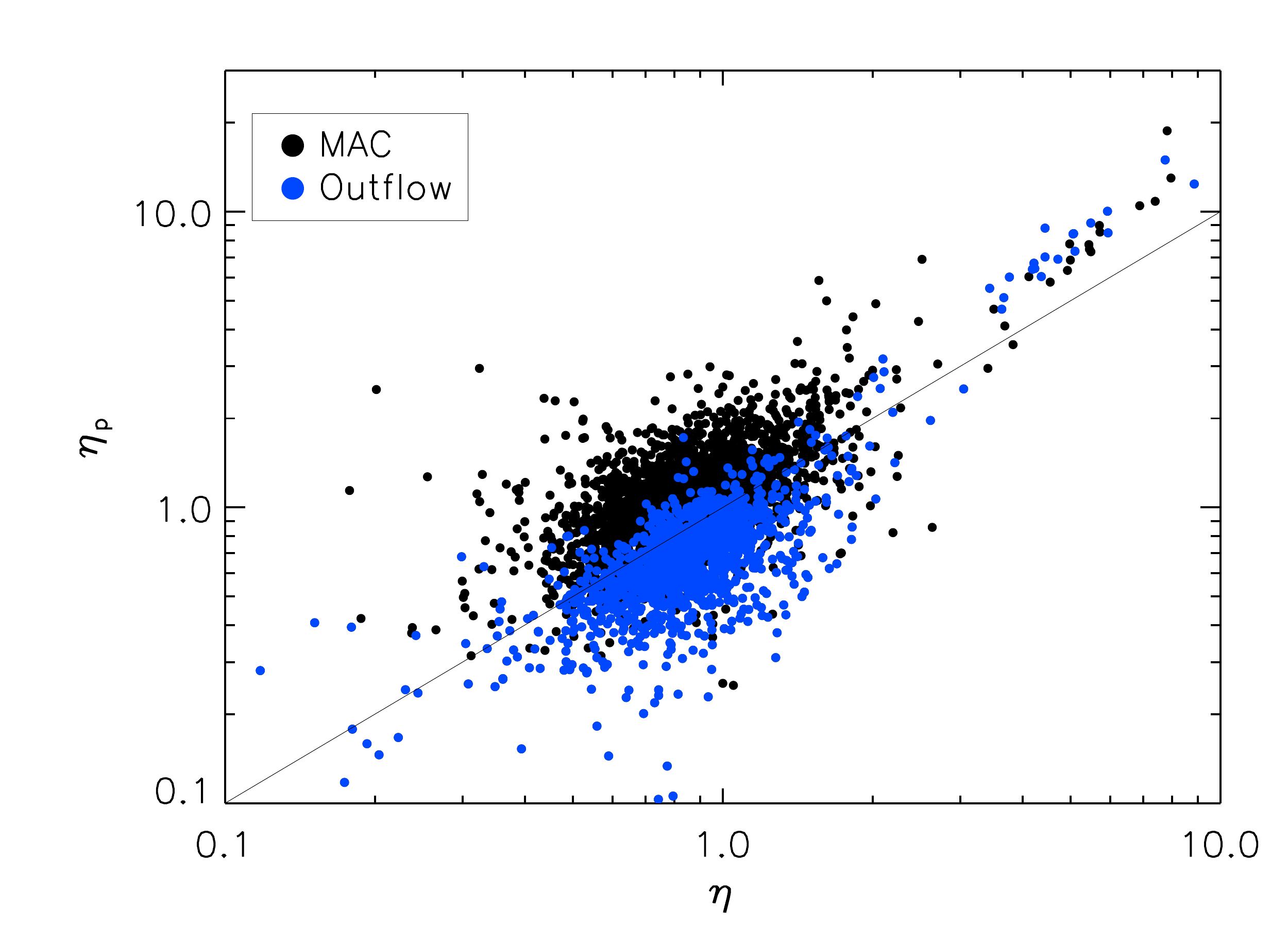}
\caption{Predicted values of $\eta_p$ compared to the value of $\eta$ measured in the FiBY simulation.  The predictions were based on three models.  If the halo has a net inflow of baryons inside $\Rcut$ compared to the previous snapshot, then we applied the MAC model (black points, Equation \ref{Equ:MAC}).  If there is a net outflow, then we applied the outflow model (blue points, Equation \ref{Equ:Pontzen}).  On top of both of these models, we then include the (minimal) effects of dynamical friction due to infalling collisionless clumps of stars and DM (Equation \ref{Eq:Lackner}).}
\label{Fig:PEta}
\end{center}
\end{figure}

  We show the comparison between $\eta_p$ and $\eta$ in Figure \ref{Fig:PEta}, where haloes to which we applied the MAC model are in black, and blue denote haloes where we used the outflow model. We find that the haloes with baryonic inflows show a decent agreement with the measurements from the simulation.  However, there is a large spread in values and a tendency to over-predict the enhancement by a factor of a few.  In contrast, the haloes with outflows tend to under-predict the enhancement.  Both of these offsets can be explained by noting that more than one physical effect may happen between snapshots which is not captured in our simple predictions.  A halo which has a net outflow of baryons may have first been adiabatically contracted before the blowout, and so the outflow model would tend to over-estimate the drop in DM, leading to too small values of $\eta_p$ for those haloes.
  
\begin{figure*}
\begin{center}
\includegraphics[width = \textwidth]{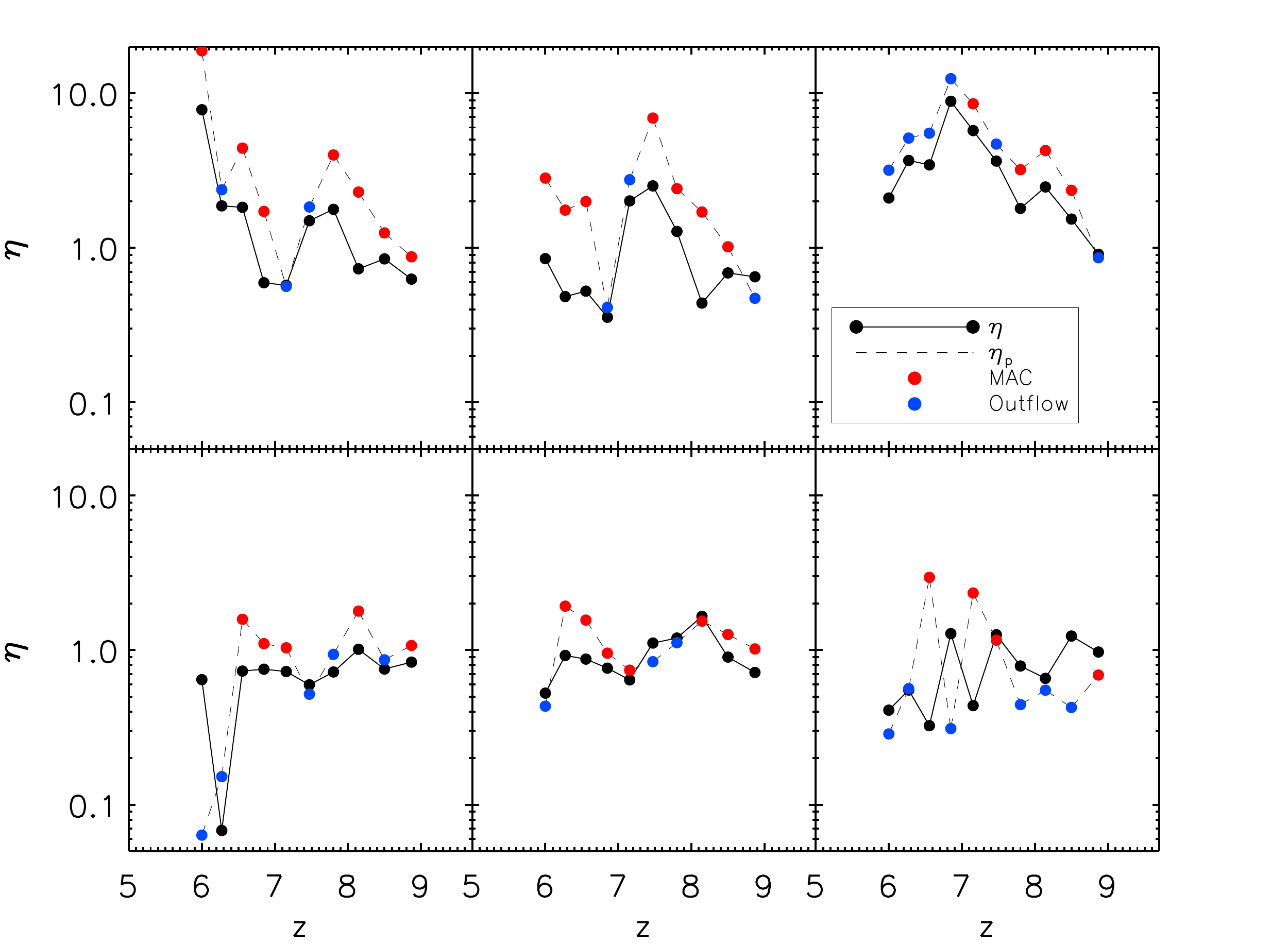}
\caption{Evolution of $\eta$ from $z=6-9$ for FiBY haloes (solid curve) and from the same models as Figure \ref{Fig:PEta} (dashed curve).  The red points show snapshots where we applied the MAC model, and the blue points show when we applied the outflow model.  The models are able to follow the trend of $\eta$ with redshift, though the MAC model tends to over-predict $\eta$ as was seen in Figure \ref{Fig:PEta}.}
\label{Fig:PEtaEvo}
\end{center}
\end{figure*}

In Figure \ref{Fig:PEtaEvo}, we show the evolution of $\eta_p$ for the same six haloes shown in Figure \ref{Fig:SFREvo} between $z=6-9$.  The solid curve shows the evolution of $\eta$ measured in the simulation, while the dashed curve shows $\eta_p$.  The points are coloured based whether or not we applied the MAC model or the outflow model at that snapshot.  The physical models capture the cyclical nature of the evolution of $\eta$ very well.  The MAC model over-predicts $\eta$ slightly as noted earlier in Section \ref{Sec:MAC}.  As the dynamical friction model barely effects the mass inside $\Rcut$, we conclude that adiabatic contraction and rapid outflows together dominate the shape of the DM profile in the centre of our haloes. 
  
\section{Discussion}
\label{sec:Disc}

Cool dense gas at the cores of haloes seems to nearly universally develop a cuspy DM profile \adb{during the early epochs studied in our simulation}.  If the gas is unable to cool and collapse to the centre, whether via reionization or stellar feedback, then the halo does not develop an enhancement of DM in the centre compared to the DMO halo.  Once an enhancement has been developed, we have shown that it is possible to destroy it via a rapid expulsion of the baryons from stellar feedback.   

We find that the star formation cycle is the main driver of the evolution of the central DM profile.  In these dwarf galaxies at high redshifts, star formation is very cyclical due to the large amount of infalling gas as these haloes are growing rapidly.  Thus, after gas is blown out via supernovae explosions, more gas is able to cool and collapse to the centres of the haloes in short time frames, starting the SF cycle again.   

\adb{We find that in the early Universe, a wide range of central halo profiles is possible.  We note that the ubiquitous cores found in \citet{Governato12} and \citet{Teyssier13} are formed after $\geq 1 \mathrm{\,Gyr}$, longer than the total cosmic time of our simulation.  During the early phases of galaxy evolution seen in our simulation, the next infall cycle brings in more mass than was blown out previously, due to the high accretion rates at early times \citep{Dekel09}.  The model of \citet{Pontzen12}  depicts a flatting of the profile even if the potential returns to its previous state before the blowout.  However, if the potential is actually deeper than its previous state, then the expected final energy of the DM can be lower than their initial energy (see Appendix \ref{Sec:PontzenDeriv}).  Hence when the mass accreted is larger than the mass blown out by feedback, such as in the early epochs of galaxy formation, then rapid potential changes do not ubiquitously lead to flattened cored profiles.}

The stochastic nature of \adb{gas inflow rates, and hence }star formation \adb{and feedback} in these high-redshift galaxies means that the DM profile \adb{in the early Universe} is not universal.  One would expect that the time scales for star formation and stellar feedback will govern the timing of the cycles of the DM profile \adb{during this early epoch}.  If the timescale for star formation is shorter than that of feedback, then gas is still flowing in, and $\eta$ should increase.  Similarly, if feedback is dominating, then that timescale will determine the drop in $\eta$.  

\begin{figure}
\begin{center}
\includegraphics[width=\columnwidth]{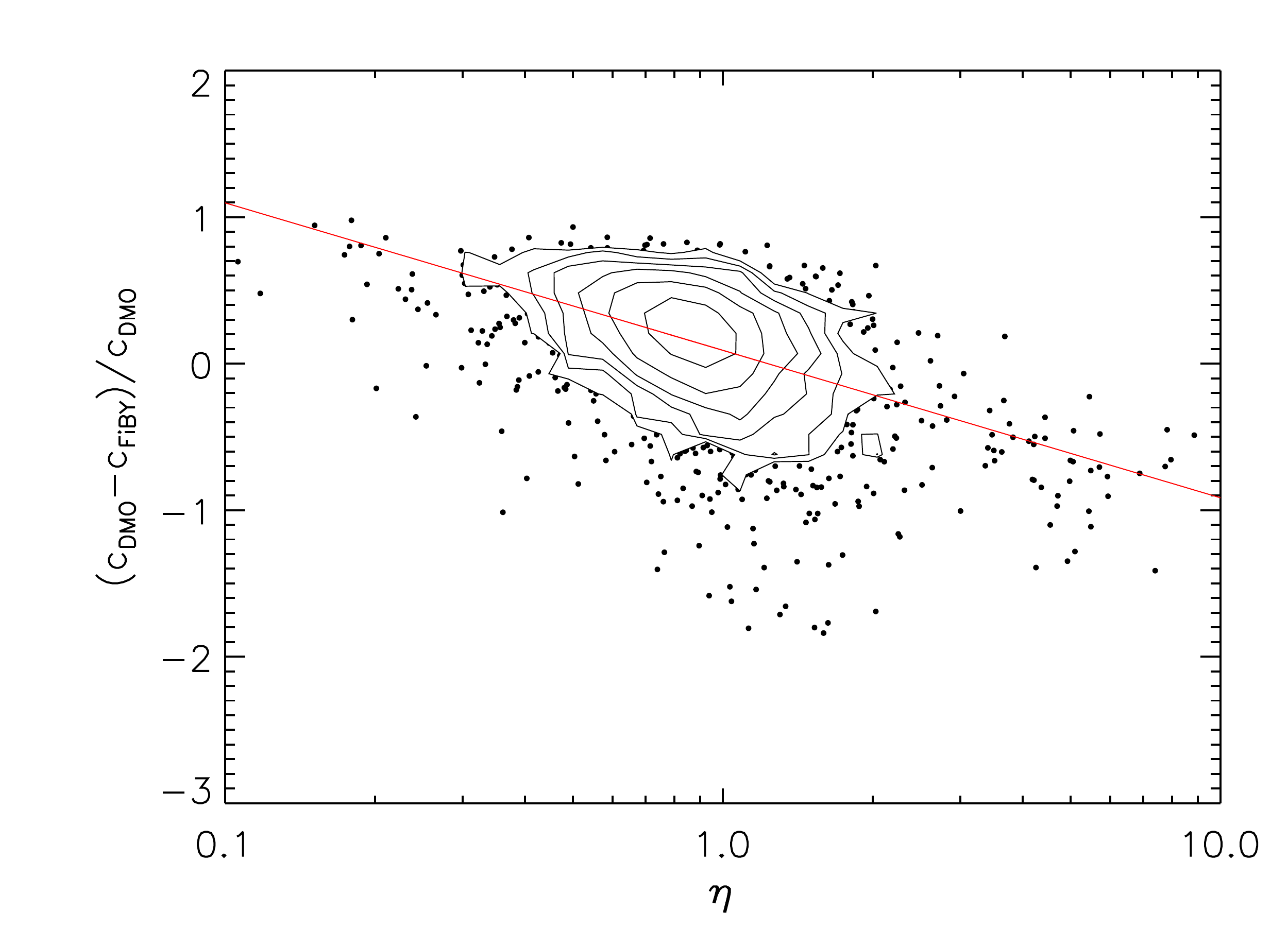}
\caption{Fractional difference in concentration between the DMO and FiBY runs as a function of $\eta$.  The solid line is a linear fit to the data.  Halos with strong enhancement show increased concentration compared to the DMO run.  The points are haloes which lie outside the lowest contour.  Halos with a strong decrement ($\eta < 1$) show smaller concentrations than their DMO counterparts.  Since enhancement is tied to the star formation cycle, the halo concentration will also be dependent on the star formation history of the halo.}
\label{Fig:DC}
\end{center}
\end{figure}

An implication of the non-universality of the central DM density profile is that of a systematic scatter in a concentration mass relation due to the presence of baryons \adb{at high redshift}.  We show in Figure \ref{Fig:DC} the change in concentration between the DMO and FiBY runs as a function of $\eta$.  Unsurprisingly, there is a systematic trend to larger concentrations with increasing $\eta$.  Since we have shown that haloes with high central SFR also show strong enhancements in the central DM, we would predict that galaxies which show strong nuclear starbursts would also show increased concentration parameters at a given halo mass, and a cuspy central DM profile.  From our simulation, we would estimate that a halo with SFR $\geq 0.1 \Msun \rm{yr}^{-1}$ inside the central $\approx 100 \pc$ will have a cuspy profile.  

The physical insights gained from our simulations may help us understand better the DM profiles of local dwarfs.  For an old stellar population which formed before or during reionization, we can assume that there have been no major gas inflows since that time.   The combined effects of stellar feedback from the last star formation episode and photo-heating from reionization imply that \adb{even at $z=6$,} the halo \adb{can} have a flattened cored profile with $\eta \leq 1$, similar to the lower mass sample (in red) in Figure \ref{Fig:Redshift}. 

\section{Conclusions}

Using a high resolution cosmological simulation of the first galaxies, we have studied the DM profiles for a large sample of haloes.  We have focused on the integrated profile in central $100\pc$ of the halo.  We compare the total mass inside $\Rcut$ on a halo to halo basis in the simulation with baryons to a dark matter only simulation of the same region.  Our main conclusions can be summarized as follows.

\begin{enumerate}
\item \adb{There is no universal inner DM profile in our haloes \adb{at these early epochs}, due to the baryonic modifications to the DM.  The distribution of $\eta$ has a mean of $\bar{\eta} = 0.94$ and a median value of $0.87$. There is a large spread of $\eta$ about the mean, and $\eta$ can vary widely during the evolution of the halo.  The distribution of $\eta$ is roughly log-normal, though at the highest halo masses, there is a significant tail with high values of $\eta$.}
\item The MAC model provides a good model for haloes with significant amounts of gas and which have $\eta > 1$.
\item Reionization inhibits the baryonic growth of haloes which cannot self-shield against the radiation. Hence $\eta$ cannot be very large for these haloes.  We find that haloes with $M \leq 3 \times 10^8 \Msun$ during reionization have smaller baryon fractions even in the core of the galaxy.  show strong baryonic quenching.
\item The evolution of the central DM is coupled to the star formation cycle of the galaxy \adb{during the first billion years studied in this work}.  As gas falls in, the DM contracts adiabatically.  However, once stellar feedback blows the gas out of the galaxy, the DM responds to the shallower potential and a drop in $\eta$ is observed which is roughly consistent with the blowout model in \citet{Pontzen12}.
\item Dynamical friction does not play a major role in removing mass from the central part of the halo, due to low stellar masses available to fall into the halo.

\end{enumerate}

\section*{Acknowledgements}
These simulations were run using the facilities of the Rechenzentrum Garching.  CDV acknowledges support by Marie Curie Reintegration Grant PERG06-GA-2009-256573.
\bibliography{AP}
\bibliographystyle{mn2e}

\appendix
\section{Adiabatic contraction derivation}
We show here the derivation of equation \ref{Equ:MACFit}, the relationship between the quality of fit of the AC model, given by $\frac{M^{\rm{FiBY}}(<r_f)}{M^{\rm{DMO}}(<r_i)}$ and $\eta$, This assumes a power law relation between the baryon ratio $\fb$ and $\eta$ of equation $\fb+1 = A\eta^B$.  Starting from equation {Equ:SAC}, and setting $M^{\rm{b}}_i = 0$ and assuming equation \ref{Equ:assume} is valid, we divide by the mass in the DMO simulation inside $\Rcut$ to obtain
\begin{equation}
\eta r_i =\left( \eta + M^{\rm{b}}(<r_f)/M^{\rm{DMO}}(<r_f) \right) r_f ,
\label{Equ:MACfb1}
\end{equation}
where we have replaced $M^{\rm{FiBY}}(<r_f)/M^{\rm{DMO}}(<r_f)$ with $\eta$.  Then, if we assume the DMO halo can be represented as a singular isothermal sphere, then we set 
\begin{equation}
\frac{M^{\rm{DMO}}(<r_f)}{r_f} = \frac{M^{\rm{DMO}}(<r_i)}{r_i}.
\label{Equ:SIS}
\end{equation}
Substituting equation \ref{Equ:SIS} into equation \ref{Equ:MACfb1}, and replacing the baryonic mass $M^{\rm{b}}(<r_f) = \fb M^{\rm{FiBY}}(<r_f)$, we obtain
\begin{equation}
\eta \frac{r_i}{r_f} = \eta + \frac{\fb M^{\rm{FiBY}}(<r_f)}{M^{\rm{DMO}}(<r_i)} \frac{r_i}{r_f}.
\end{equation}
Using equation \ref{Equ:SAC} again, we can write the ratio of initial to final radii as $1+\fb$.  Substituting this then gives
\begin{equation}
\frac{\eta}{1+\fb} = \frac{M^{\rm{FiBY}}(<r_f)}{M^{\rm{DMO}}(<r_i)}.
\end{equation}
Lastly, inserting the observed power law relationship between $1+\fb$ and $\eta$, we predict a relationship between the quality of fit of the MAC model, as given by the ratio $\frac{M^{\rm{FiBY}}(<r_f)}{M^{\rm{DMO}}(<r_i)}$, and $\eta$, such that
\begin{equation}
\frac{M^{\rm{FiBY}}(<r_f)}{M^{\rm{DMO}}(<r_i)} = \frac{1}{A} \eta^{1-B}.
\end{equation}

\section{Instantaneous blowout}
\label{Sec:PontzenDeriv}
\adb{We show here how the model of \citet{Pontzen12} does not always imply a net energy gain, if the potential becomes deeper than it was initially due to additional inflow of baryons via cosmic accretion.  We follow the simple analytic model in their Section 3, that of a 1D spherically symmetric power law potential:
\begin{equation}
V(r;t) = V_{\mathrm{0}} r^n.
\end{equation}
We assume that the normalization constant, $V_{\mathrm{0}}$ is only a function of time, and not space.   If we allow the potential to change such that $V_{\mathrm{1}} = V_{\mathrm{0}} + \Delta V_{\mathrm{0}}$, then the expected value for the energy after the change is $E_{\mathrm{1}} = E_{\mathrm{0}} + \Delta E_{\mathrm{1}}$, where the expected value of the change in energy is given by their equation 3:
\begin{equation}
\left< \Delta E_{\mathrm{1}} \right> = \frac{2 E_{\mathrm{0}}}{2+n} \frac{\Delta V_{\mathrm{0}}}{V_{\mathrm{0}}}.
\end{equation}
If the potential switches immediately by some fraction $\alpha$ of its initial change, then we can calculate the total shift in energy as follows.  Letting $\Delta V_{\mathrm{1}} = - \alpha \Delta V_{\mathrm{0}}$, then the fraction $\alpha$ propagates through as a simple pre-factor to equation 5 of \citet{Pontzen12}.  Taylor expanding and simplifying, we get the final total energy as
\begin{equation}
\left<E_f\right> \simeq E_{\mathrm{0}} + (1-\alpha) \frac{2 E_{\mathrm{0}}}{2+n}\frac{\Delta V_{\mathrm{0}}}{V_{\mathrm{0}}} + \alpha \frac{2nE_{\mathrm{0}}}{(2+n)^2} \left( \frac{\Delta V_{\mathrm{0}}}{V_{\mathrm{0}}} \right)^2.
\end{equation}
In the case where $n < 0$, $E_{\mathrm{0}} < 0$ for bound orbits and hence the third term is always positive.  In cases where $\alpha = 1$ then the final total energy is always larger than the initial energy.  However, when $\alpha > 1$, the second term is negative (assuming $V_{\mathrm{0}} < 0$) and can dominate the change in energy, leading to a system more bound than its initial configuration.  This requires 
\begin{equation}
\frac{\alpha - 1}{\alpha} > \frac{n}{2+n} \left(\frac{\Delta V_{\mathrm{0}}}{V_{\mathrm{0}}}\right).
\end{equation}
For the Keplerian case of $n=-1$, and assuming that $\frac{\Delta V_{\mathrm{0}}}{V_{\mathrm{0}}} = -0.1$, we find that $\alpha > 1.11$ to have a net drop in expected value of the final energy.  
}

\end{document}